\documentclass[titlepage,12pt]{article}
\usepackage{amssymb,epsfig,pslatex}
\usepackage{cite}
\usepackage{float}
\usepackage{wrapfig}
\usepackage{hhline}
\usepackage{dcolumn}
\restylefloat{figure}

\makeatletter
\def\@sect#1#2#3#4#5#6[#7]#8{\ifnum #2>\c@secnumdepth
  \def\@svsec{}\else
  \refstepcounter{#1}\edef\@svsec{\csname the#1\endcsname.\hskip0.5em}\fi
  \@tempskipa #5\relax
  \ifdim \@tempskipa>\z@
    \begingroup
      #6\relax
      \@hangfrom{\hskip #3\relax\@svsec}{\interlinepenalty \@M #8\par}%
    \endgroup
    \csname #1mark\endcsname{#7}\addcontentsline
      {toc}{#1}{\ifnum #2>\c@secnumdepth \else
        \protect\numberline{\csname the#1\endcsname}\fi #7}%
  \else
    \def\@svsechd{#6\hskip #3\@svsec #8\csname #1mark\endcsname
      {#7}\addcontentsline{toc}{#1}{\ifnum #2>\c@secnumdepth \else
        \protect\numberline{\csname the#1\endcsname}\fi #7}}%
  \fi \@xsect{#5}}

\def\lsim{\mathrel{\raise.3ex\hbox{$<$\kern-.75em\lower1ex\hbox{$\sim$}}}}
\def\gsim{\mathrel{\raise.3ex\hbox{$>$\kern-.75em\lower1ex\hbox{$\sim$}}}}

\def\gev{\rm GeV}
\def\msb{\overline{\rm MS}}
\def\rpv{\not\!\!{R} }

\usepackage{axodraw}

\begin{document}

\begin{titlepage}

\begin{center}
{\LARGE {\bf 
QCD corrections to single slepton production\\  at hadron colliders}}
\\
\vspace{2cm}
{\large{\bf
Yu-Qi~Chen$^1$, Tao Han$^{1,2,3}$, Zong-Guo~Si$^4$, }}
\par\vspace{1cm}
 $^1$ Institute of Theoretical Physics, Academia Sinica,
           Beijing 100080, China\\
 $^2$ Department of Physics, University of Wisconsin,
               Madison, WI 53706, USA\\
 $^3$  Department of Physics, Center for High-energy Physics Research,
Tsinghua University, Beijing 100080,  China\\
 $^4$  Department of Physics, Shandong University,
Jinan Shandong 250100, China \\ \vspace{1cm}
E-Mail: {\tt ychen@itp.ac.cn}, {\tt than@physics.wisc.edu}, 
{\tt zgsi@sdu.edu.cn}
\par\vspace{1cm}

{\bf Abstract}\\
\parbox[t]{\textwidth}
{
We evaluate the cross section for single slepton production at hadron
colliders in supersymmetric theories with $R$-parity violating interactions
to the next-to-leading order in QCD.
We obtain fully differential cross section by using the phase space
slicing method. We also perform soft-gluon resummation to all
order  in $\alpha_s$ of  leading logarithm to obtain a complete 
transverse momentum spectrum of the slepton. 
We find that  the full transverse momentum 
spectrum is peaked at a few  GeV, consistent with the early
results for Drell-Yan production of lepton pairs.
We also consider the contribution from gluon fusion
via quark-triangle loop diagrams dominated by the $b$-quark loop. 
The cross section of this process is significantly smaller
than that of the  tree-level process induced by the initial $b\bar{b}$
annihilation.
}

\end{center}
\vspace*{2cm}
                                                                                

\end{titlepage}

\section{Introduction}

The recent evidence of neutrino oscillation unambiguously indicates
non-zero masses of neutrinos \cite{Nureview}.
This  poses a fundamental question:
Whether or not the color and electrically neutral neutrinos are
Majorana fermions, whose mass term violates the lepton number
by two units. In general,  lepton-number violating interactions
of $\Delta L=1,2$ will induce a Majorana mass for neutrinos.
Thus searching for  the effects of
lepton-number violation is of fundamental importance to establish the
Majorana nature of the neutrinos and of significant implications for
particle physics, nuclear physics and cosmology.

Weak scale supersymmetry (SUSY) remains to be a leading candidate for
physics beyond the standard model (SM). While SUSY models often yield
 rich physics in the flavor sector, theories with  $R$-parity violation may
have lepton-number violating interactions  and thus provide a
Majorana mass  for neutrinos \cite{Rmass,Rnew}. Proposals for searching
for slepton production via $R$-parity violating interactions
at hadron colliders already exist in the literature \cite{Rold}.
Experimental searches are being actively conducted at the Tevatron collider.
Despite of the negative results \cite{tevrp},  it is believed that it is conceivable
to observe a signal of single slepton production at the
Tevatron and the LHC if the interaction strengths is sufficiently large
 and is suitable for the neutrino mass generation \cite{Rnew,Barbier04ez}.

In this paper,
we evaluate the cross section for a single slepton production in
supersymmetric theories with $R$-parity violating interactions
at the LHC to the next-to-leading order in QCD. 
We present  fully differential cross sections  by using the phase space
slicing method.  Our results, 
when overlapping, agree with an earlier calculation \cite{Choudhury2003}.
We also  perform soft-gluon resummation to all order
at leading log to obtain a complete transverse momentum spectrum.
Our results here are consistent with a recent calculation \cite{Yang:2005ts}.
We further  consider the contribution with gluon fusion
via quark-triangle diagrams, which could help probe the
$R$-parity violating couplings involving $b$ quarks.
We find that the cross section for this process is  smaller than
the  process  induced by  initial $b\bar{b}$ state.

The paper is organized as follows:
The cross sections of all single slepton production processes 
at leading order and next-to-leading order
are  presented in Sections \ref{sec2} and \ref{sec3} respectively,
in a pedagogical manner.
Numerical results at the Tevatron and the LHC  are  given and
discussed in Section \ref{sec3}. 
The transverse momentum distribution of a
slepton is calculated in Section \ref{sec4}. We further investigate the slepton
production via gluon fusion through a heavy-quark loop in Section \ref{sec5}.
Finally a short summary is given in Section \ref{sec6}.

\section{Born Level results}
\label{sec2}

\begin{figure}[tb]
\begin{center}
\psfig{figure=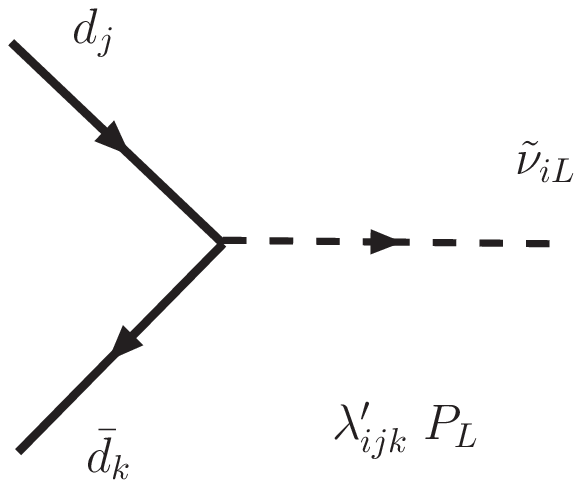,width=3.67cm}
\psfig{figure=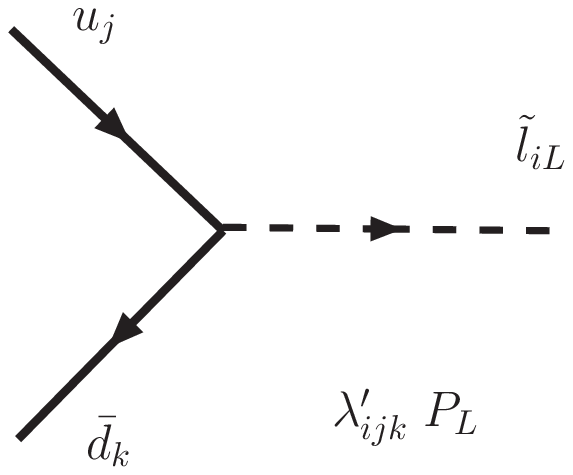,width=3.67cm}
\psfig{figure=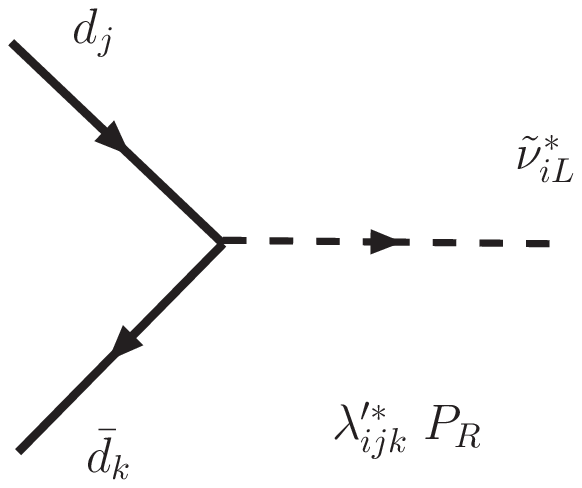,width=3.67cm}
\psfig{figure=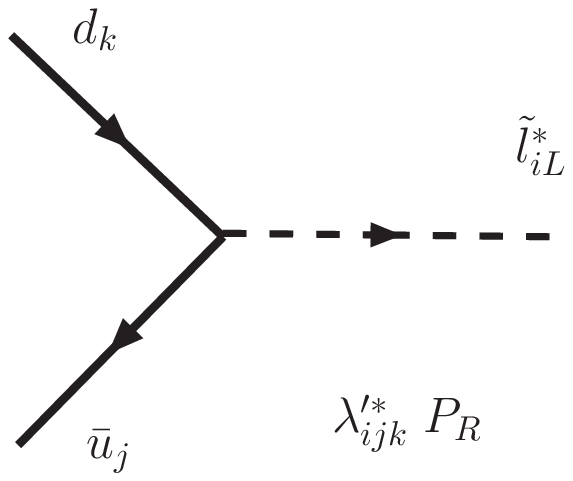,width=3.67cm}
\caption{\small  Feynman diagram for
 $\bar{q}'(p_1)+q(p_2)\rightarrow \tilde{\ell}(q)$ in the $\not\!\!{R} $  SUSY
 model.}
\label{figborn}
\end{center}
\end{figure}


In this section, we set up our notation and the initial results for the
Born-level cross sections. 
In $\not\!\!{R}$ SUSY models with a $\lambda'$ coupling as shown
in Fig.~\ref{figborn}, a  slepton or a  sneutrino
can be produced via quark anti-quark annihilation. The interaction
couplings are proportional to $\lambda'_{ijk}$ and are flavor-dependent.
At the leading order,
neglecting the flavor index, the generic process is
\begin{equation}
\bar{q}(p_1)+q'(p_2)\rightarrow \tilde{\ell}(q),
\label{proc-qqsl}
\end{equation}
where $p_1$, $p_2$ and $q$ denote the four momenta of the
corresponding particles. The invariant amplitude for this process reads
\[ {\cal M}^{(0)} = \left\{
\begin{array}{ll}
-i\,\lambda'\, \bar{v}(p_1)\,P_L\, u(p_2) &  \mbox{for}\ \tilde{\nu},\ \tilde{\ell}, \\
-i\,\lambda'^*\, \bar{v}(p_1)\,P_R\, u(p_2)  &  \mbox{for}\ \tilde{\nu}^*,\ \tilde{\ell}^*.
\end{array}
\right. \]
%
with $P^{}_{R,L}  = (1\pm \gamma_5)/2$. The Born-level cross section,
after averaging over the initial spins and colors, is thus given by
\begin{equation}
\sigma_{Born}(\hat{s})= \frac{1}{N_c}\, \frac{1}{2\hat{s}}\,
\int \frac{d^{d-1}q}{(2\pi)^{d-1}2q_0} \, \frac{1}{4}\,|{\cal
M}^{(0)}|^2 \,(2\pi)^d \, \delta^d(p_1+p_2-q)
={ {{\lambda'}^2 \pi} \over {4N_c \hat{s} }} \delta(1-\tau),
\end{equation}
where $\hat{s}=(p_1+p_2)^2,\ \tau\,=\,m_{\tilde{\ell}}^2/\hat{s}$,
and $d=4-2\epsilon$ is set to be 4 for the Born-level calculations.

\section{Next-to-leading order QCD corrections }
\label{sec3}

The leading order processes (\ref{proc-qqsl}) receive QCD
corrections.  At the next leading order, they may arise from a
real gluon emitting process, from one-loop virtual corrections, or
from the gluon initiated process. As  is well-known, although infrared
(IR) divergences appear in each diagram individually, they cancel
by adding them together. There are also collinear divergences
arising from the fermion mass singularities in the real gluon
emission and gluon splitting processes.
They can be absorbed to the redefinition of the
parton distribution functions. The ultraviolet (UV) divergences
arising from the virtual corrections can be removed by
renormalization of the coupling constants. Throughout this paper,
we adopt dimensional regularization to regulate both the
infrared and the ultraviolet divergences and use $\msb$
scheme to carry out the renormalization and mass factorization.

\subsection{Virtual corrections }

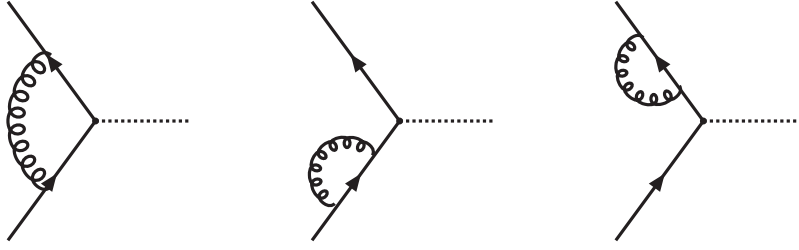
\begin{figure}[tb]
\begin{center}
\begin{picture}(100,100)(0,0)
\SetWidth{1.2}
\Vertex(35,50){1.3}
\ArrowLine(2,5)(35,50)
\ArrowLine(35,50)(2,95)
\DashLine(35,50)(70,50){1.2}
\GlueArc(35,50)(30,124,236){3}{8}
\end{picture}~~~~
\begin{picture}(100,100)(0,0)
\SetWidth{1.2}
\Vertex(35,50){1.3}
\ArrowLine(2,5)(35,50)
\ArrowLine(35,50)(2,95)
\DashLine(35,50)(70,50){1.2}
\GlueArc(15,30)(12,36,248){2}{7}
\end{picture}~~~~
\begin{picture}(100,100)(0,0)
\SetWidth{1.2}
\Vertex(35,50){1.3}
\ArrowLine(2,5)(35,50)
\ArrowLine(35,50)(2,95)
\DashLine(35,50)(70,50){1.2}
\GlueArc(16,70)(12,108,326){2}{7}
\end{picture} \\
\caption{\small  Feynman diagrams for virtual gluon corrections to
 $\bar{q}(p_1)+q(p_2)\rightarrow \tilde{\ell}(q)$.}
 \label{fignlov}
\end{center}
\end{figure}

We first calculate the contributions from the virtual corrections
as shown in Fig.~\ref{fignlov}.  With a straightforward calculation,
we obtain the contribution to the cross section from the vertex diagram as
%
\begin{eqnarray}
\label{s-vertex}
&& \sigma_{vertex}(\hat{s}) =  \frac{C_F\alpha_s}{\pi} C_{\epsilon}
\,\Big( \,-\frac{1}{\epsilon^2}+\frac{\pi^2}{2}-1\,\Big) \sigma_{Born}(\hat{s})
\equiv K'\  \sigma_{Born}(\hat{s}) \\
&& C_{\epsilon}= \frac{1}{\Gamma(1-\epsilon)}
 \,\Big(\,\frac{4\pi\mu^2}{\hat{s}}\,\Big)^{\epsilon},\ \
C_F= {N_c^2-1 \over 2 N_c}.
\end{eqnarray}
Note that the single $1/\epsilon$ pole is absent due
to the cancellation between the IR and the UV poles.

The contribution from the one loop self-energy diagrams vanishes,
since there is no energy scale involved for an on-shell massless quark,
thus the virtual corrections are simply given by the vertex correction.
Adding the contributions of the Born term and the virtual gluon corrections
together, the cross section is
\begin{eqnarray}
\sigma_{Born}\,+\,\sigma_{virtue}
 & = &
 {{\lambda'}^2\,\pi \over 4 N_c \,\hat{s} }\;\delta(1\,-\,\tau) \;
 \left (1 + K' \right)\; \nonumber\\
& \approx &
 {{\lambda'}_R^2\,\pi \over 4 N_c\,\hat{s}}\;\delta(1\,-\,\tau) \;
 \left [\;
 (\,Z_{\lambda}\,Z_2^{-1}\,)^2 + K' \right]\;,
 \label{sigma-0+v}
\end{eqnarray}
where  ${\lambda'}_R$ is the renormalized coupling constant,
$Z_{{\lambda}}$ and $Z_2$ are the vertex and wave function
renormalization constants  extracted from the ultraviolet poles
in the one-loop vertex and self-energy diagrams in Fig.~\ref{fignlov},
respectively, and are given in the ${\msb}$ scheme by
\begin{eqnarray}
&& Z_{{\lambda}} =
1 - \frac{\alpha_s}{4\pi\Gamma(1-\epsilon)}\ C_F
(4\pi)^{\epsilon}\ \frac{4}{\epsilon}, \label{Z_lambda}\quad
Z_2 =  1 - \frac{\alpha_s}{4\pi\Gamma(1-\epsilon)}\ C_F
(4\pi)^{\epsilon}\  \frac{1}{\epsilon} ,\\
&& {\lambda'}_R = {\lambda'}\, Z_{\lambda}\,Z_2^{-1} \approx
 1 - \frac{\alpha_s}{4\pi\Gamma(1-\epsilon)}\ C_F
(4\pi)^{\epsilon}\  \frac{3}{\epsilon} \label{Z2} .
\end{eqnarray}

Combining the results above,
we obtain the cross section with the
Next-to-Leading Order (NLO) virtual corrections are
\begin{eqnarray}
\sigma_{Born}+\sigma_{virtue} =
 {{\lambda'}_R^2 \pi \over 4 N_c \hat{s}} \delta(1-\tau)
 \left [1 + \frac{C_F \alpha_s}{\pi} C_{\epsilon} \Big(
-\frac{1}{\epsilon^2}-\frac{3}{2\epsilon}
-\frac{3}{2}\ln\frac{\hat{s}}{\mu^2} +\frac{\pi^2}{2}-1 \Big) \right]. ~
 \label{S-0-V}
\end{eqnarray}
The  infrared and collinear  singularities
are isolated by the double and single poles in $\epsilon$.

\subsection{Real gluon emission }
The real gluon emission  processes
%
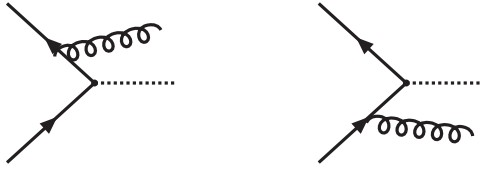
\begin{figure}[tb]
\begin{center}
\begin{picture}(100,100)(0,0)
\SetWidth{1.2}
\Vertex(35,50){1.3}
\ArrowLine(2,20)(35,50)
\ArrowLine(35,50)(2,80)
\DashLine(35,50)(65,50){1.2}
\Gluon(20,60)(60,70){3}{5}
\end{picture}~~~~~
\begin{picture}(100,100)(0,0)
\SetWidth{1.2}
\Vertex(35,50){1.3}
\ArrowLine(2,20)(35,50)
\ArrowLine(35,50)(2,80)
\DashLine(35,50)(65,50){1.2}
\Gluon(20,35)(60,30){3}{5}
\end{picture} \\
\caption{\small  Feynman diagrams for
 $\bar{q}'(p_1)+q(p_2)\rightarrow \tilde{l}(q)+g(k)$
in the $\not\!\!{R} $  SUSY model.}
\label{fignloqq}
\end{center}
\end{figure}
\begin{equation}
\bar{q}(p_1)+q(p_2)\rightarrow g(k)+\tilde{l}(q)\;, \label{qqgs}
\end{equation}
as shown in Fig.~\ref{fignloqq}  also contribute  to the
next leading order QCD corrections.
The scattering matrix element and its square are
\begin{eqnarray}
{\cal M}^a & = &
 -i {\lambda'}  g_s\mu^\epsilon \,
\epsilon^{\mu}(k) \bar{v}(p_1) \,T^a\,\Big(\, \gamma_{\mu}
\frac{\not{p}_1 - \not{k}}{2p_1\cdot k} \, \frac{1+\gamma_5}{2} \,
- \, \frac{1+\gamma_5}{2} \, \frac{\not{p}_2-\not{k}}{2p_2\cdot
k}\, \gamma_{\mu} \Big)\, u(p_2),
\nonumber \\
|{\cal M}^a|^2 & = & 16
{\lambda'}^2\,g_s^2\mu^{2\epsilon}\,N_c\,C_F\,{1 \over \cos ^2
\theta}
 \,\left(
 {1 \over (1-\tau )^2} -{1\over 1-\tau } + { 1-\epsilon \over 2}
 \right)\;,
 \label{M2-qqgs}
\end{eqnarray}
where $\theta $ is the angle between the outgoing gluon and the
initial anti-quark in the parton-level center mass frame system. The
Lorentz invariant phase space integral reads:
\begin{eqnarray}
\int\, d\, PS_2^{\epsilon} &\;=\;& \int
\frac{d^{3-2\epsilon}k}{(2\pi)^{3-2\epsilon}\,2E_g}\,
\frac{d^{3-2\epsilon}q}{(2\pi)^{3-2\epsilon}\,2q_0}\,
\delta^{4-2\epsilon} (p_1+p_2-q-k) \nonumber \\
&\;=\;&{(4\pi)^{-1+\epsilon} \over \Gamma(1-\epsilon)} \;
{(1-\tau)^{1-2\epsilon} \over 2 {\hat s}^\epsilon }
 \; \int ^\pi_0
\, d\theta \, \sin ^{1-2\epsilon} \theta \;.
 \label{LIPS}
\end{eqnarray}
In order to retain kinematical distributions, we wish not to integrate
out  $\theta$.  We adopt the phase space
slicing method \cite{Harris01sx} to separate the IR and the collinear
divergences. These singularities are isolated into the soft and
collinear regions by two parameters $x_{min}$ and $\delta$,
respectively, according to
\begin{eqnarray}
&& E_g < \frac{\sqrt{\hat{s}}}{2}x_{min},
 \label{Soft}\\
&& 1-|\cos(\theta)| < \delta .
 \label{Coll}
\end{eqnarray}

\subsubsection{Soft region}
In the soft region as defined by the gluon energy in Eq.~(\ref{Soft}),
the contribution to the cross section is given by
\begin{eqnarray}
\sigma_{soft}&\;=\;&\frac{1}{4N_c^2}\, \frac{1}{2\hat{s}}\,
\int\,d\, PS_2^\epsilon \, |M^a|^2
\,\Theta(\frac{\sqrt{\hat{s}}}{2}x_{min}-E_g)\,
 \nonumber \\
&\;=\;&  \sigma_{Born}(\hat{s})\
\frac{C_F\alpha_s}{\pi} \,C_{\epsilon} \left(\frac{1}{\epsilon^2} -\frac{2\ln x_{min}}
{\epsilon}+2\ln^2
x_{min}-\frac{\pi^2}{6}\,\right),
 \label{S-soft}
\end{eqnarray}
We see that the above IR double pole just cancels that of the
virtual correction in (\ref{S-0-V}) as expected. The single pole
is artificial and arises from the explicit cutoff.

\subsubsection{Collinear region}
The collinear region is constrained by Eq.~(\ref{Coll}) but in order to
avoid double-counting, one must impose $E_g >\frac{\sqrt{\hat{s}}}{2}x_{min}$.
Its contribution to the cross section is given by
\begin{eqnarray}
\sigma_{col}&=&\frac{1}{4N_c^2}\, \frac{1}{2\hat{s}}\,
\int\,d\, PS_2\, |M^{a}|^2
\,\Theta(E_g-\frac{\sqrt{\hat{s}}}{2}x_{min})\, \,\Theta(\delta-
1+|\cos(\theta)|)\,
\nonumber \\
&=& \sigma_{Born}(x \hat{s}) \
\frac{C_F\alpha_s}{\pi} \,C_{\epsilon} \,
\left(-\frac{2^{\epsilon}}{\epsilon}\,\right)\, (1-\epsilon
\ln\delta) \int\limits_0^{1-x_{min}}\, dx\,
\frac{1+x^2-\epsilon(1-x)^2}{(1-x)^{1+2\epsilon}}, \label{sig-col}
\end{eqnarray}
where $1-x$ is the gluon energy-momentum fraction with respect to the
parent quark defined as
$k=(1-x) p_1 ~{\rm or}~k=(1-x) p_2,$ depending on which quark the
gluon is radiated from.
%

With the help of the ``plus function",
\begin{equation}\label{plus}
\Big[F(x)\Big]_+=\lim\limits_{\beta\rightarrow 0} \,\Big[\,
\Theta(1-x-\beta)\, F(x)\,-\,\delta(1-x-\beta)\,
\int\limits_0^{1-\beta} F(y) d y\,\Big]\,,
\end{equation}
Eq.~(\ref{sig-col}) can be rewritten as
\begin{eqnarray}
\sigma_{col}&=&\frac{C_F\alpha_s}{\pi} C_{\epsilon}\ \Big\{
 \sigma_{Born}(\hat{s})\Big[\frac{2\ln
x_{min}}{\epsilon}-2\ln^2x_{min} -2\ln x_{min} \ln
\frac{\delta}{2}\Big] \\
&+&
 \int\limits_0^1 dx ~\sigma_{Born}(x\hat{s})
\Big[\left(-\frac{1}{\epsilon}+\ln \frac{\delta}{2} \right)
\frac{1+x^2}{(1-x)_+} +(1-x)
+2 (1+x^2)\Big(\frac{\ln(1-x)}{1-x}\Big)_+\Big] \Big\} .
\nonumber
\end{eqnarray}
Adding the contributions from leading term, virtual corrections,
soft part, and the collinear part together, we have:
\begin{eqnarray}
&& \sigma_{Born}\,+\,\sigma_{virtue}\,+\,\sigma_{soft}\,+\,
\sigma_{col}
  \nonumber \\
&=&
 {{\lambda'}_R^2\,\pi \over 4 N_c\,\hat{s}}\;\delta(1\,-\,\tau)
 \left[1 +  \frac{C_F \alpha_s}{\pi}\, C_{\epsilon}\, \Big(
 -\frac{3}{2\epsilon}-\frac{3}{2}\ln\frac{\hat{s}}{\mu^2}
 +\frac{\pi^2}{3}-1-2 \ln x_{min}\ln{\delta \over 2}
  \Big)\,\right]
  \label{Sig-0vsc} \\
   &+&
 \frac{C_F\alpha_s}{\pi} C_{\epsilon}
 \int\limits_0^1 dx ~\sigma_{Born}(x\hat{s})
 \left[\Big(\ln \frac{\delta}{2}-\frac{1}{\epsilon} \Big)
 \frac{1+x^2}{(1-x)_+} +(1-x) +2
 (1+x^2)\Big(\frac{\ln(1-x)}{1-x}\Big)_+\right] .   \nonumber
\end{eqnarray}
 The collinear divergence in $1/\epsilon$ is due to the
massless approximation of the incoming quarks, and can be
absorbed into the definition of  parton densities.
Within $\msb$ scheme, the universal counter terms for the
collinear subtraction are given by
\begin{eqnarray}
\sigma_{col}^{CT}&=&\frac{C_F \alpha_s}{2\pi}
\frac{1}{\Gamma(1-\epsilon)} \Big(\frac{4\pi
\mu^2}{\mu_F^2}\Big)^{\epsilon} \Big\{
\frac{3}{\epsilon}\sigma_{Born}(\hat{s})
+\frac{2}{\epsilon}\int\limits_0^1 dx
\sigma_{Born}(x\hat{s})\frac{1+x^2}{(1-x)_+}\Big\}\nonumber \\
&=&\frac{C_F \alpha_s}{\pi}
\Big[\frac{C_{\epsilon}}{\epsilon}+\ln\frac{\hat{s}}{\mu_F^2}\Big]\,
\Big[\frac{3}{2}\sigma_{Born}(\hat{s}) +\int d x
\,\frac{1+x^2}{(1-x)_+} \sigma_{Born}(x \hat{s}) \,\Big],
\label{Sig-ct}
\end{eqnarray}
where $\mu_F$ is the factorization scale. Adding (\ref{Sig-0vsc}) and
(\ref{Sig-ct}) together and integrating over
$\sigma_{Born}(x\hat s)= {\lambda'}_R^2\,\pi /( 4 N_c\,\hat{s})  \delta(x - \tau)$,
we obtain
\begin{eqnarray}
&& \sigma_{Born}\,+\,\sigma_{virtue}\,+\,\sigma_{soft}\,+\,
\sigma_{col}\,+\, \sigma_{col}^{CT}
  \nonumber \\
& = &
 {{\lambda'}_R^2\,\pi \over 4 N_c\,\hat{s}}\;\delta(1\,-\,\tau) \;
 \left[\;1 \;+\;
 \frac{C_F \alpha_s}{\pi}\, \Big(\,
 -\frac{3}{2}\ln\frac{\mu^2_F}{\mu^2}\,
 +\,\frac{\pi^2}{3}-1-2 \ln x_{min}\ln{\delta \over 2} \,
  \Big)\,\right]\;
   \nonumber \\
   & + &
 \frac{C_F\alpha_s}{\pi} \
  {{\lambda'}_R^2\,\pi \over 4 N_c\,\hat{s}}
 \left[\, (\ln \frac{\delta}{2}+\ln{\hat s\over \mu^2_F} )
 \frac{1+\tau^2}{(1-\tau)_+} +(1-\tau) +2
 (1+\tau^2)\,\Big(\,\frac{\ln(1-\tau)}{1-\tau}\Big)_+\,\right] . ~ \quad
 \label{Sig-0vsc-ct}
\end{eqnarray}
 Obviously,  the infrared  and the
collinear divergences are cancelled as expected. We note, however,
there are still  explicit  dependences on the hard cutoff parameters
$x_{min}$ and $\delta$.
%

\subsubsection{Hard scattering}
The tree-level $2\to 2$ processes as in Eq.~(\ref{qqgs})
contribute to the order of $\alpha_s$.
The region of hard scattering is constrained by $E_g >
\frac{\sqrt{\hat{s}}}{2}x_{min} $ and $ 1-|\cos(\theta)| >\delta$.
Since the contribution to the cross section from this region is
finite, we can take $\epsilon=0$ in the calculation. In fact,
it is a regularized $2\to 2$ ``resolvable" process
and the differential cross section is written as
\begin{eqnarray}
\sigma_{2\to 2}&=& \frac{1}{4N_c^2}\, \frac{1}{2\hat{s}} \int
dPS_2 |{\cal M}^a|^2 ~\Theta(1-\delta-|\cos\theta|)
\Theta(E_g-\frac{\sqrt{\hat{s}}}{2}x_{min}) \nonumber\\
&=&\frac{C_F\alpha_s{\lambda'}^2}{4N_c\hat{s}} \int
d\cos \theta\ \Theta(1-\delta-|\cos \theta | )
\Theta(1-x_{min}-\tau) \frac{1}{1-\cos^2 \theta}\
\frac{1+\tau^2}{1-\tau}.\quad~
\label{qq222d}
\end{eqnarray}
It is Eqs.~(\ref{Sig-0vsc-ct}) and (\ref{qq222d}) that give the complete cross section
for the $q\bar q$ initiated process at the order of $\alpha_s$.

The  dependence on the cutoff parameters $x_{min}$ and $\delta$
are explicit in these two equations, reflecting the arbitrariness of our definition
of  ``soft" and ``collinear" regions. However, the summed total cross section
should be independent of those choices. To see this feature, we choose to
integrate over $\theta$ and  have
\begin{eqnarray}
\sigma_{2\to 2}
& =&-\,\frac{C_F\alpha_s{\lambda'}^2}{4N_c\hat{s}}\,
\Theta(1-x_{min}-\tau)\, \frac{1+\tau^2}{1-\tau}\,
\ln\frac{\delta}{2} \nonumber\\
&=& \frac{C_F\alpha_s{\lambda'}^2}{4N_c\hat{s}}\,\left[\,
-\ln\frac{\delta}{2}\, \frac{1+\tau^2}{(1-\tau)_+} \,+\, 2
\ln\frac{\delta}{2}\,\ln x_{min}\,\delta(1-\tau) \,\right],
\label{qq222}
\end{eqnarray}
where the ``plus-function" of Eq.(\ref{plus})
in the limit $x_{min}\rightarrow 0$
\begin{equation}
\frac{1+\tau^2}{1-\tau}
\Theta(1-x_{min}-\tau)\;=\;\frac{1+\tau^2}{(1-\tau)_+} \,-\,2 \ln
x_{min}\,\delta(1-\tau)
\end{equation}
has been used.

Putting (\ref{Sig-0vsc-ct}) and (\ref{qq222}) together, we obtain the total
cross section of $q\bar{q}$ processes to the next leading order
\begin{eqnarray}
\hat{\sigma}_{q\bar{q}}
 & = & \sigma_{Born}+\sigma_{vir}+\sigma_{col}+\sigma_{col}^{CT}+\sigma_{2\to 2} \nonumber \\
 & = & {{\lambda'}_R^2\,\pi \over 4 N_c \hat s } \,
 \left\{ \,\delta(1-\tau) \;+\;
 \frac{C_F \alpha_s}{\pi}\,
 \left[\,\left(\,
\frac{3}{2} \ln\frac{\mu^2}{\mu_F^2}+\frac{\pi^2}{3}-1 \,\right)\,
 \delta(1-\tau)\,
 \right. \right. \nonumber\\
 && \,+\, \left. \left.
  \frac{1+\tau^2}{(1-\tau)_+} \,
  \ln\frac{\hat{s}}{\mu_F^2}\,+\,(1-\tau) \,
  +\,2 (1+\tau^2) \,\left(\,\frac{\ln(1-\tau)}{1-\tau}\,\right)_+ \,
  \right]\;\right\} ,
\end{eqnarray}
where we have verified the independence of the cutoff parameters.

\subsection{Gluon-initiated process}
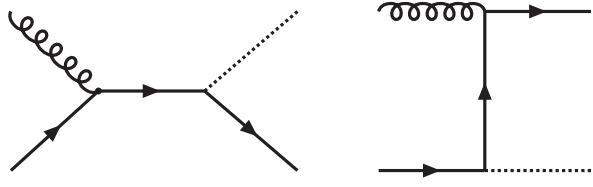
\begin{figure}[tb]
\begin{center}
\begin{picture}(100,100)(0,0)
\SetWidth{1.2}
\Vertex(35,50){1.3}
\ArrowLine(2,20)(35,50)
\Gluon(35,50)(2,80){3}{5}
\ArrowLine(35,50)(75,50)
\ArrowLine(75,50)(110,20)
\DashLine(75,50)(110,80){1.2}
\end{picture}~~~~~~
\begin{picture}(100,100)(0,0)
\SetWidth{1.2}
\ArrowLine(20,20)(60,20)
\Gluon(20,80)(60,80){3}{5}
\ArrowLine(60,20)(60,80)
\ArrowLine(60,80)(100,80)
\DashLine(60,20)(100,20){1.2}
\end{picture} \\
\caption{\small  Feynman diagrams for
 $g(p_1)+q(p_2)\rightarrow \tilde{l}(q)+q(k)$ in the $\not\!\!{R} $  SUSY
 model.}
 \label{fignloqg}
\end{center}
\end{figure}

In next leading order, the gluon-initiated process
\begin{equation}
g(p_1)+q(p_2) \rightarrow \tilde{l}(q)+q(k),
\end{equation}
also contributes to the cross section. Two Feynman diagrams
contributing to the process are shown in Fig.~\ref{fignloqg}. 
With the Feynman rule given in Fig.~\ref{figborn}, 
the cross section can be written as
\begin{equation}
\tilde{\sigma}_{res}=\frac{1}{4N_c (N_c^2-1)} \frac{1}{2\hat{s}}
\int |\tilde{\cal M}|^2 ~\Theta(1-\delta-|\cos\theta|) ~
dPS_2^\epsilon
\end{equation}
where
\begin{eqnarray}
\tilde{\cal M}^a &=& -i{\lambda'} g_s \mu^\epsilon \, \epsilon^{\mu
*}(p_1) \, \bar{u}(k) \,T^a \,\Big(\, P_R \,
\frac{\not{p}_1+\not{p}_2}{\hat{s}} \gamma_{\mu} \,+\,\gamma_{\mu}
\frac{\not{p}_1-\not{k}}{2p_1\cdot k} \,P_R \, \Big)\, u(p_2) \\
|\tilde{\cal M}^a|^2 &=& {\lambda'}^2 g_s^2 N_c C_F \,\left[\,
\frac{4}{1-\cos \theta }\,
\left(\,\frac{1}{1-\tau}-2\tau\,\right)\,-\,(\,3+\cos
\theta\,)\,(\,1-\tau\,)\, \right] .
\end{eqnarray}

There are collinear divergences but no infrared divergence for
a gluon splitting to $q\bar q$. Similar to the case of $q\,\bar{q} \,\to\, \tilde{l}
\,g$ process, we can separate the contributions to the cross
section into collinear part and hard
 by imposing a cutoff
$1-|\cos\theta|=\delta $ on the phase space, where $\theta$ is the
angle between the moving direction of the initial gluon and that
of the outgoing quark.
Integrating over $\theta$ within $1-|\cos\theta|<\delta $,  we
obtain the collinear part:
\begin{eqnarray}
\tilde{\sigma}_{col}&=&-\frac{\alpha_s}{4\pi} C_{\epsilon}
\int\limits_0^1 dx ~\sigma_{Born}(x\hat{s})\,
\frac{2^{\epsilon}}{\epsilon}\, \Big(1-\epsilon \ln\delta \Big)\,
\frac{x^2+(1-x)^2}{(1-\epsilon)(1-x)^{2\epsilon}} \nonumber \\
&=&\frac{\alpha_s}{4\pi} C_{\epsilon} \
{{\lambda'}^2\,\pi \over 4 N_c \hat s } \
\Big\{1+\Big[\tau^2+(1-\tau)^2\Big]\,
\Big[-\frac{1}{\epsilon}+\ln\,\frac{\delta}{2}+2\ln(1-\tau)-1\Big]
\Big\},
\end{eqnarray}
where $x$ is defined as $k=(1-x) p_1$.

Within $\msb$ scheme, the universal collinear counter term is given by
\begin{eqnarray}
\tilde{\sigma}_{col}^{CT}&=&
\frac{\alpha_s}{4\pi} \frac{1}{\Gamma(1-\epsilon)}
\Big(\frac{4\pi\mu^2}{\mu_F^2}\Big)^{\epsilon} \frac{1}{\epsilon} \int\limits_0^1 \, d x
\sigma_{Born}(x\hat{s}) ~\Big\{x^2+(1-x)^2\Big\}  \nonumber \\
&=& \frac{\alpha_s}{4\pi}
\Big(\frac{C_{\epsilon}}{\epsilon}+\ln\frac{\hat{s}}{\mu_F^2}\Big)\,
{{\lambda'}^2\,\pi \over 4 N_c \hat s } \
\Big[\tau^2+(1-\tau)^2\Big].
\end{eqnarray}
Then we have
\begin{eqnarray}
\tilde{\sigma}_{col}+\tilde{\sigma}_{col}^{CT}
&=&\frac{ {\lambda'}^2 \alpha_s}{16N_c\hat{s}}\,\Big\{
1+\Big[\tau^2+(1-\tau)^2\Big]\,\Big[\ln \frac{\delta \ \hat{s} }{2\ \mu_F^2}
+2\ln(1-\tau)-1\Big]
\Big\} .
\end{eqnarray}

The ``resolvable" $2\to 2$ cross section is free from collinear singularity and
can be calculated in 4-dimension
\begin{eqnarray}
\tilde{\sigma}_{2\to 2}&=&\frac{{\lambda'}^2\,\alpha_s}{16 N_c \hat{s}}\,
\int\limits_{-1}^{1-\delta} \Big\{
\frac{1}{1-\cos\theta}\,\Big[(1-\tau)^2+\tau^2\Big]-\frac{3+\cos\theta}{4}\,(1-\tau)^2\Big\}\,
d\cos\theta  \nonumber \\
&=&\frac{{\lambda'}^2\,\alpha_s}{16 N_c \hat{s}}\,
\Big\{ -\Big[(1-\tau)^2+\tau^2\Big]\,\ln\frac{\delta}{2}
-\frac{3}{2}\,(1-\tau)^2\Big\}.
\end{eqnarray}
The final result of the total cross section is independent of the cutoff
parameters and  is given by
\begin{eqnarray}
\hat{\sigma}_{gq}&=&\hat{\sigma}_{g\bar{q}}
 =\tilde{\sigma}_{col}+\tilde{\sigma}_{col}^{CT}+\tilde{\sigma}_{2\to 2}
 \nonumber \\
&=& \frac{{\lambda'}^2 \alpha_s}{16N_c\hat{s}}\,\left\{
\Big[\tau^2+(1-\tau)^2\Big]\,\Big[\ln \frac{\hat{s}}{\mu_F^2}
+2\ln(1-\tau)\Big]+\frac{(1-\tau)(7\tau-3)}{2} \right\}.
\end{eqnarray}
We find that our analytical results at NLO QCD are in agreement
with those in the literature \cite{Choudhury2003}.

\subsection{Total cross sections at hadron colliders}

Given the leading and the next leading order cross sections of the
subprocesses  above,  we now investigate the total
cross sections of the slepton production at hadron colliders such
as the Tevatron and the LHC  at the next leading order.

The total cross sections are formally given
by the QCD factorization formula as
%
%
%
\begin{eqnarray}
 \sigma(S) = \sum_{i,j} \int dx_1 dx_2 \; f_i(x_1,\mu)\,
 f_{j}(x_2,\mu) \hat{\sigma}_{ij}(x_1 x_2 S,\mu)  ,
 \label{QCD-fac}
\end{eqnarray}
where $f_{i,j}$  are parton distribution functions (PDF) and $i,j$
sum over all possible types of partons contributing to the subprocesses,
$\mu$ is the factorization scale. In our numerical calculations,
we consistently take to the leading and the next leading order parton
distribution functions, CTEQ6L and CTEQ6.1M parton distribution
functions \cite{Pumplin02vw} for the leading order for the next leading order
calculations, respectively and take $\mu$ to be the slepton mass.

A useful quantity to describe the contributions from the next
leading order corrections is the $K$-factor which is defined as
the ratio of the next leading total cross section to the leading
order one, $ K \equiv \sigma_{NLO}/ {\sigma_{LO}} $. We will
show the numerical results of $ {\sigma_{LO}}$, $ \sigma_{NLO}$,
and the corresponding $K$-factor for various processes at the
Tevatron and the LHC energies.

\begin{figure}[tb]
\centerline{\psfig{figure=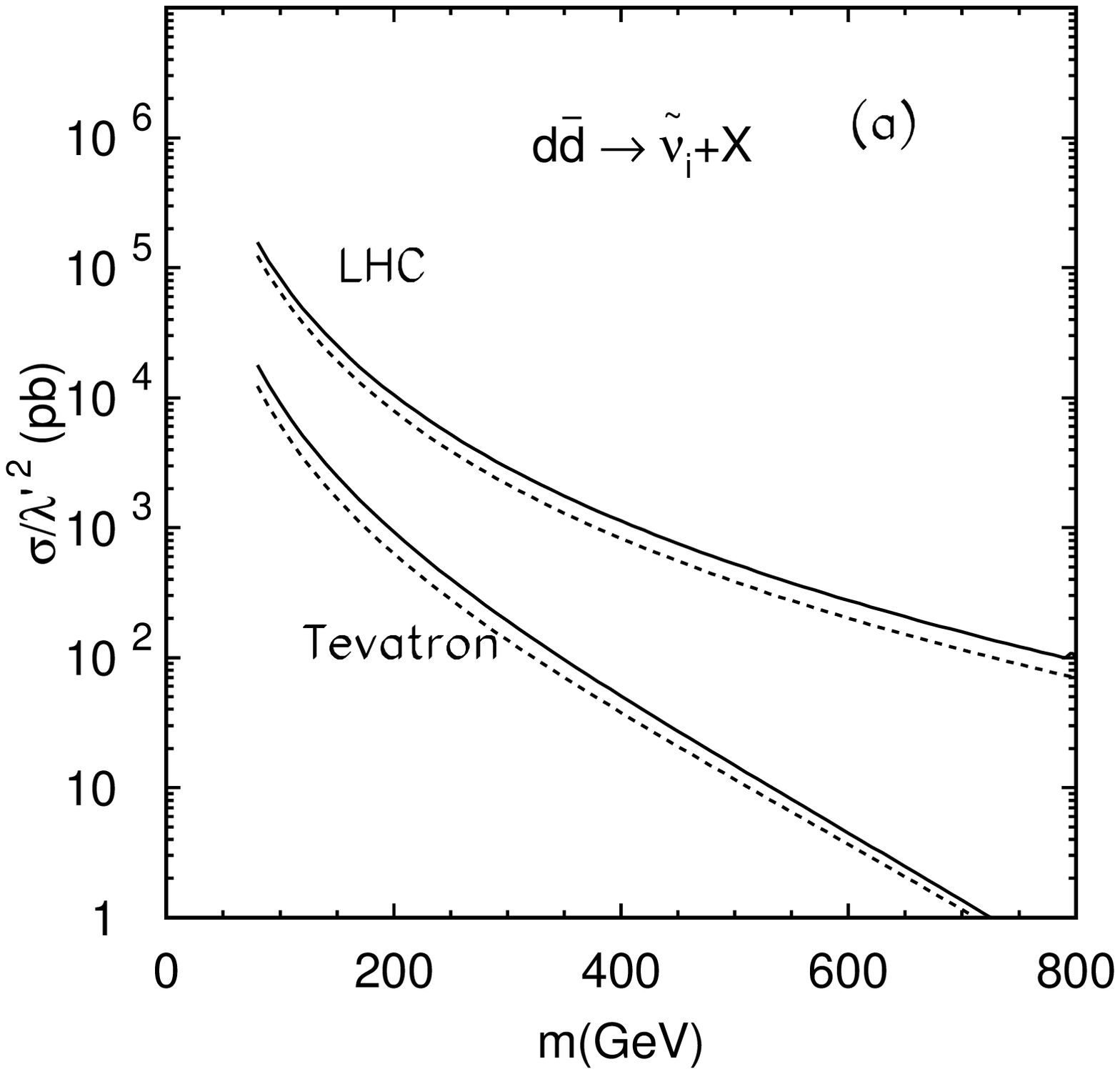,width=8cm}
                    \psfig{figure=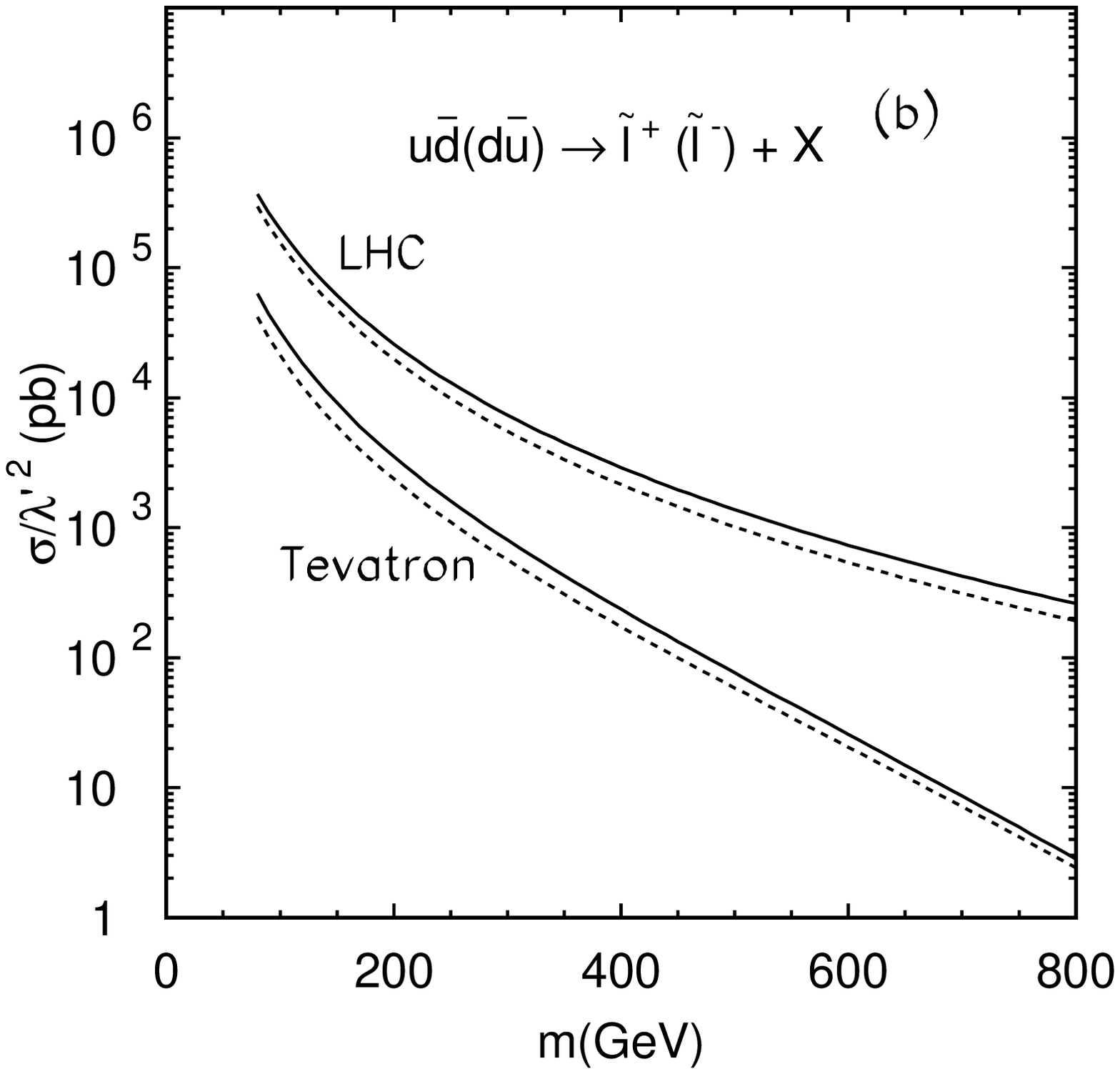,width=8cm}}
\centerline{\psfig{figure=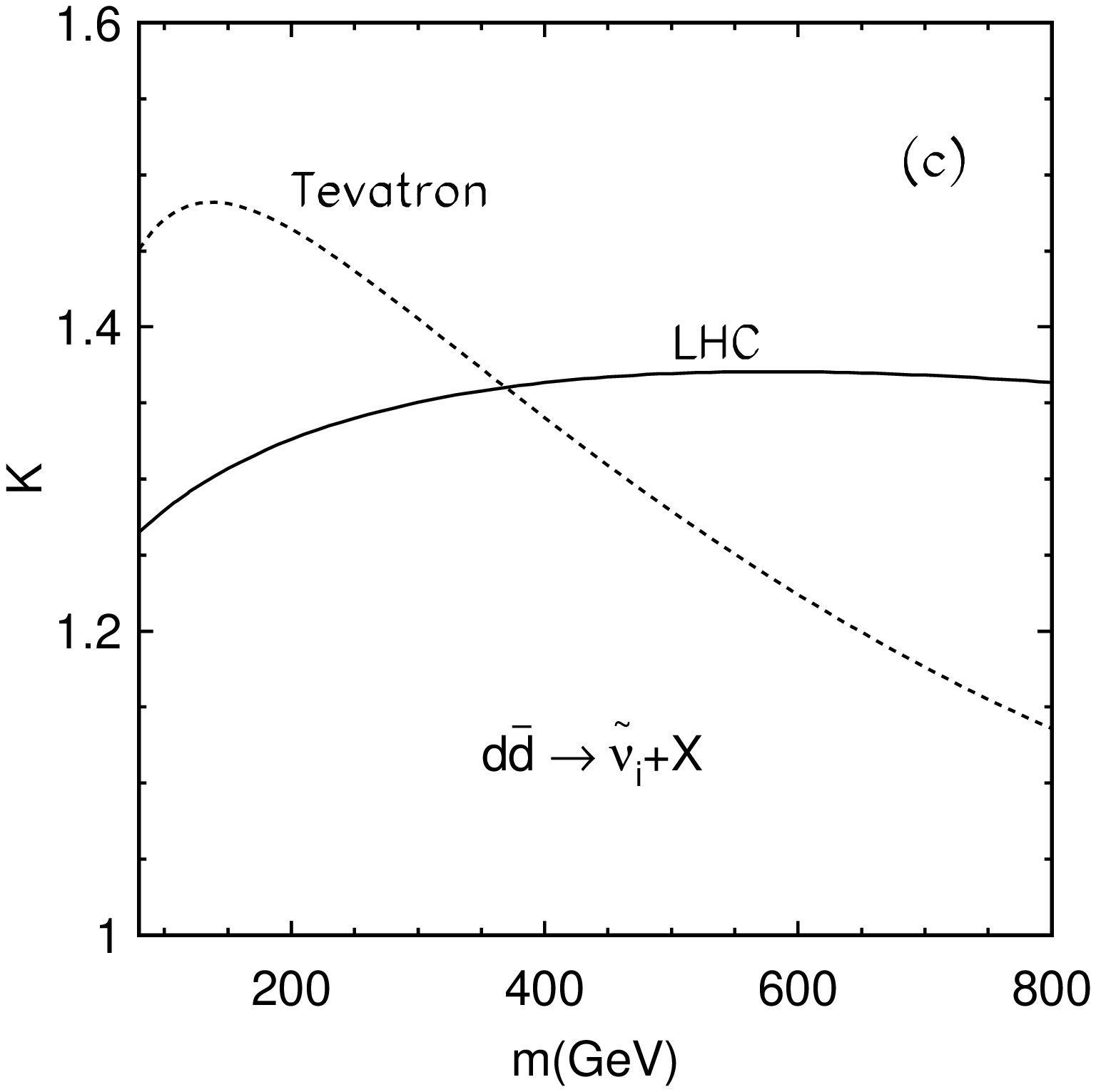,width=8cm}
                     \psfig{figure=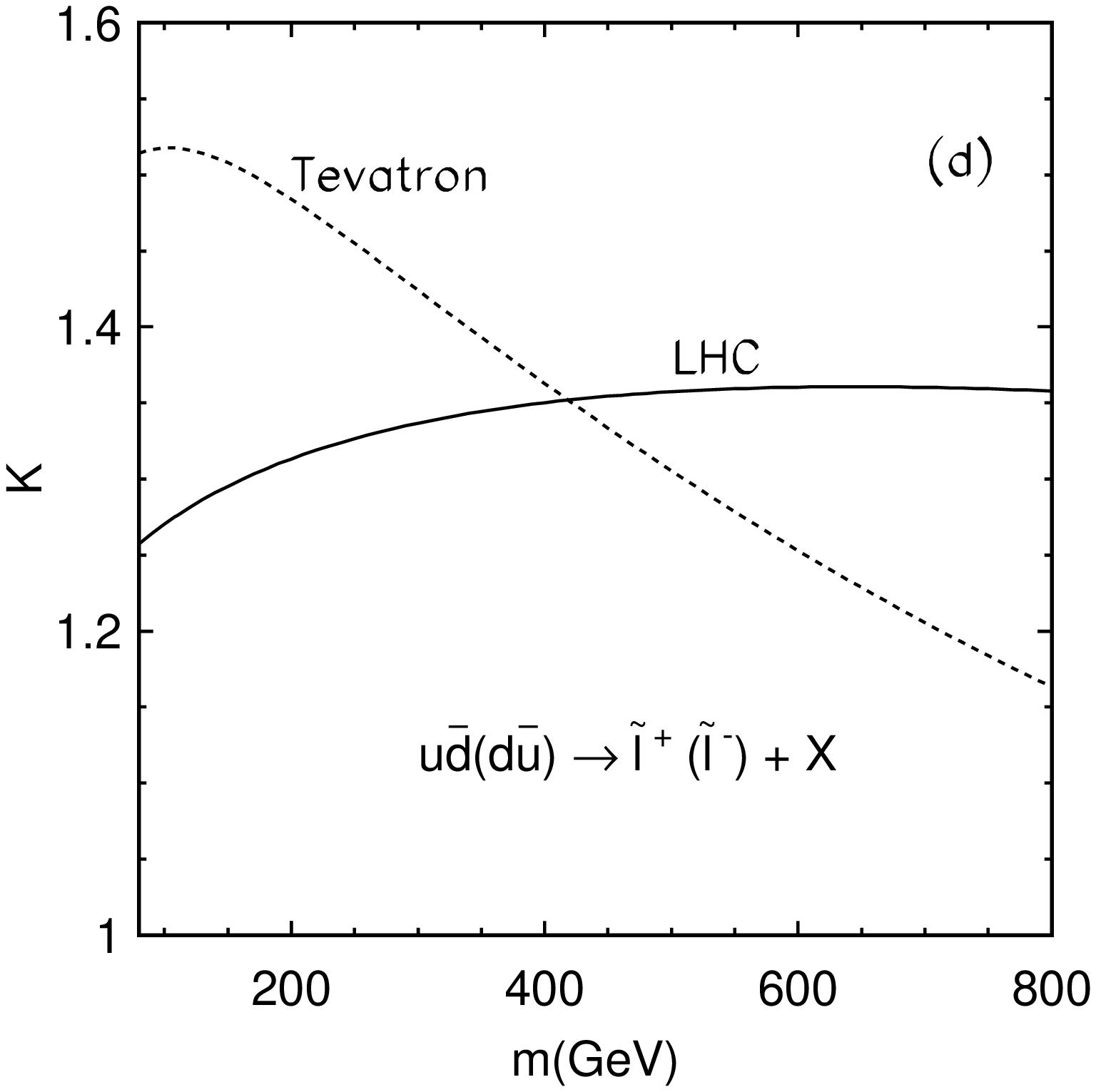,width=8cm}}
\caption{Cross sections at the LO (dashed line) and the NLO (solid line) and their
corresponding $K$-factor related to ${\lambda'}_{i11}$ for $d\bar{d}\rightarrow \tilde{\nu}_i+X$
shown in (a) and (c),  and $u\bar{d}(d\bar{u})\rightarrow
\tilde{\ell}^+(\tilde{\ell}^-)$ in (b) and (d) at the Tevatron ($\sqrt{s}=2$ TeV) and
the LHC ($\sqrt{s}=14$ TeV). 
}\label{csnlol11}
\end{figure}

\begin{figure}[tb]
\centerline{\psfig{figure=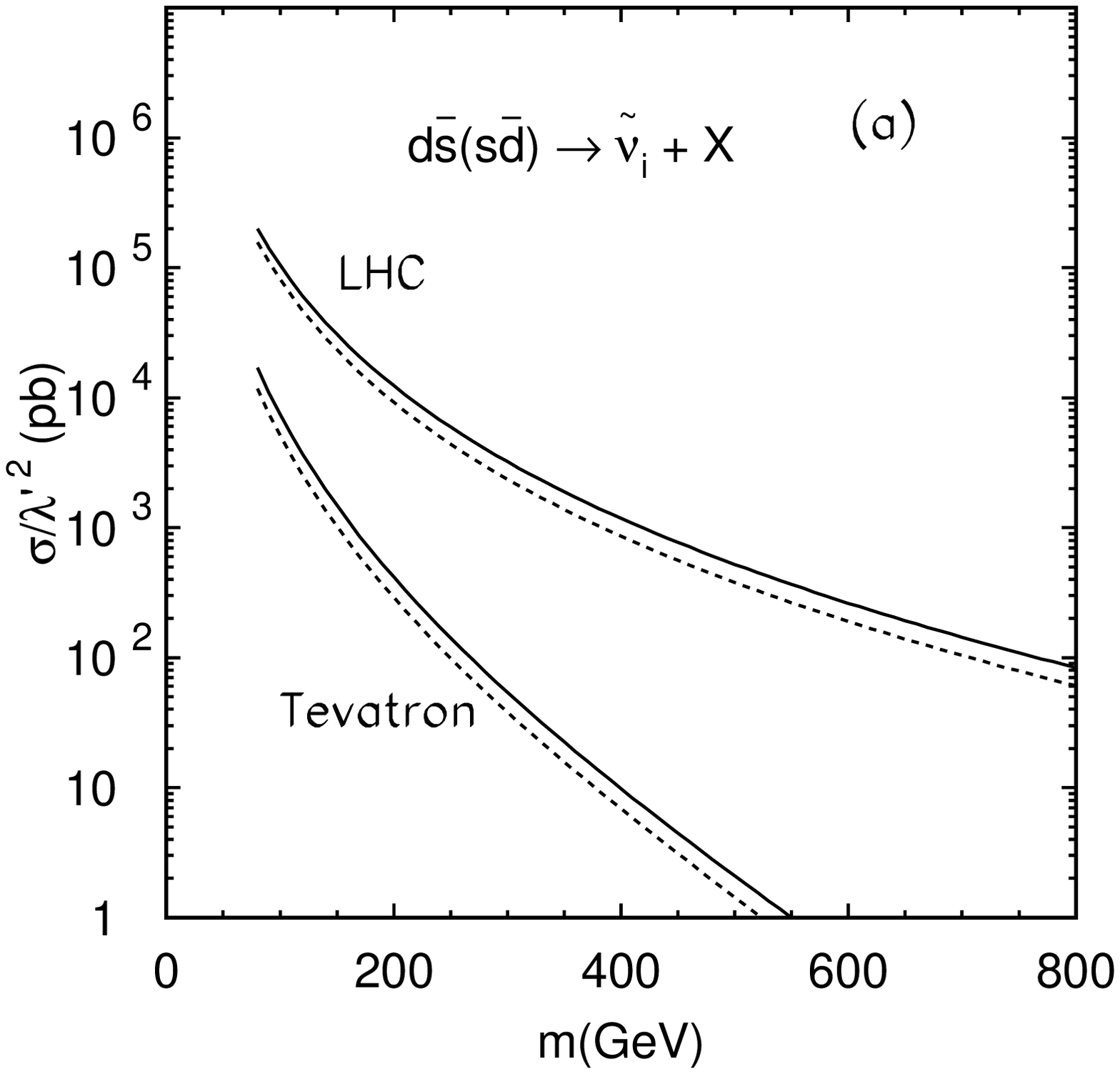,width=8cm}
\psfig{figure=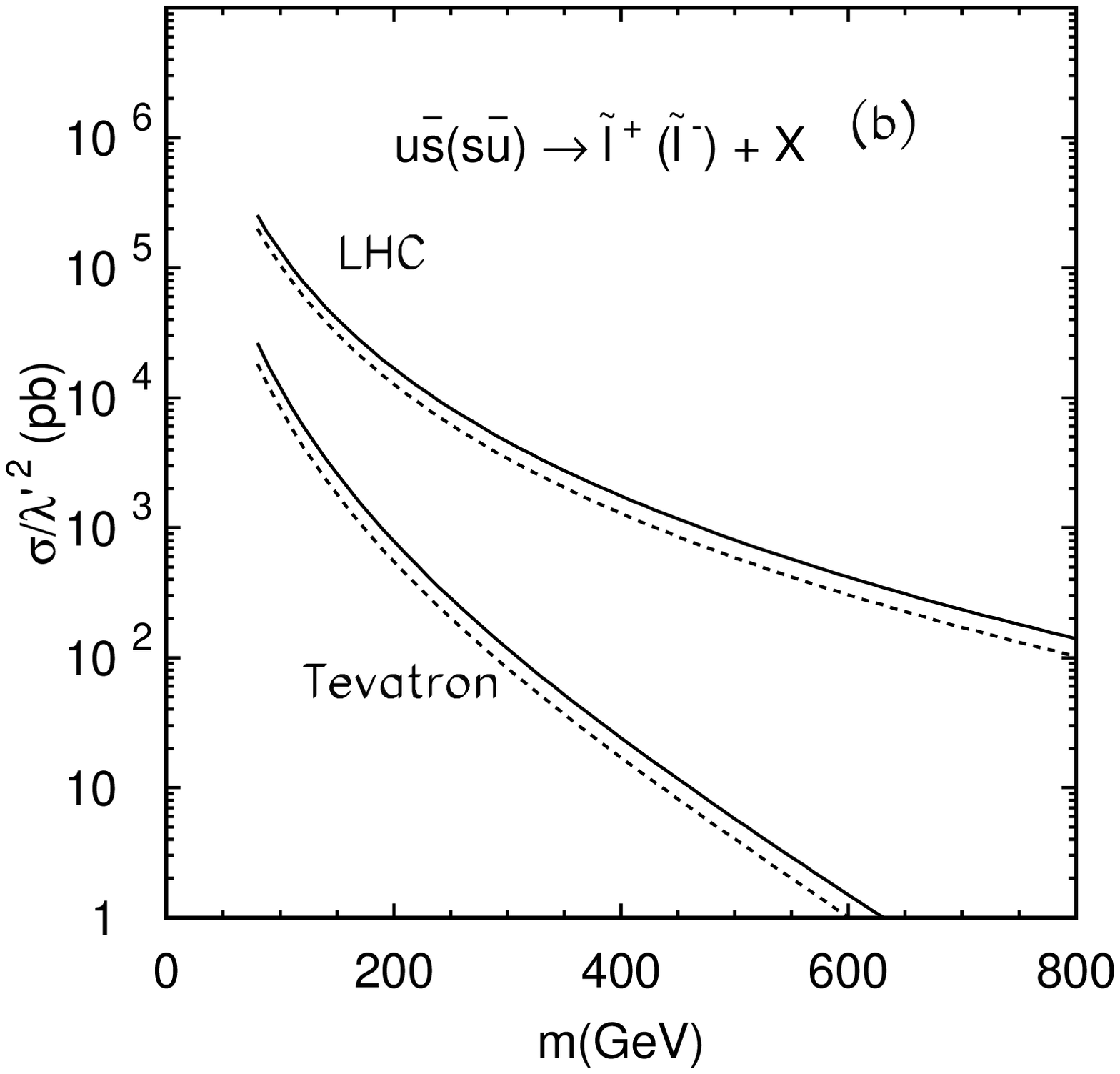,width=8cm}}
\centerline{\psfig{figure=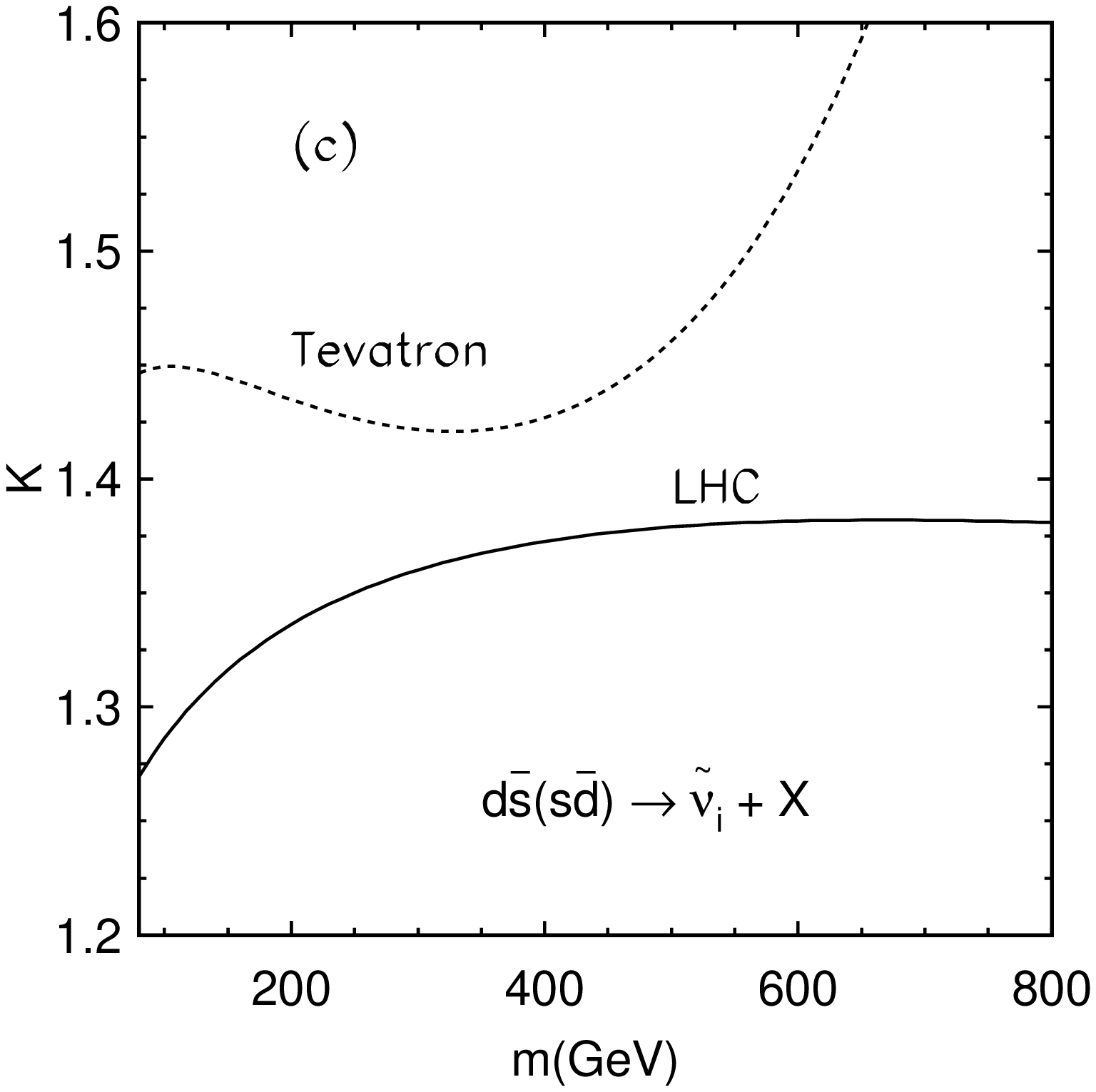,width=8cm}
\psfig{figure=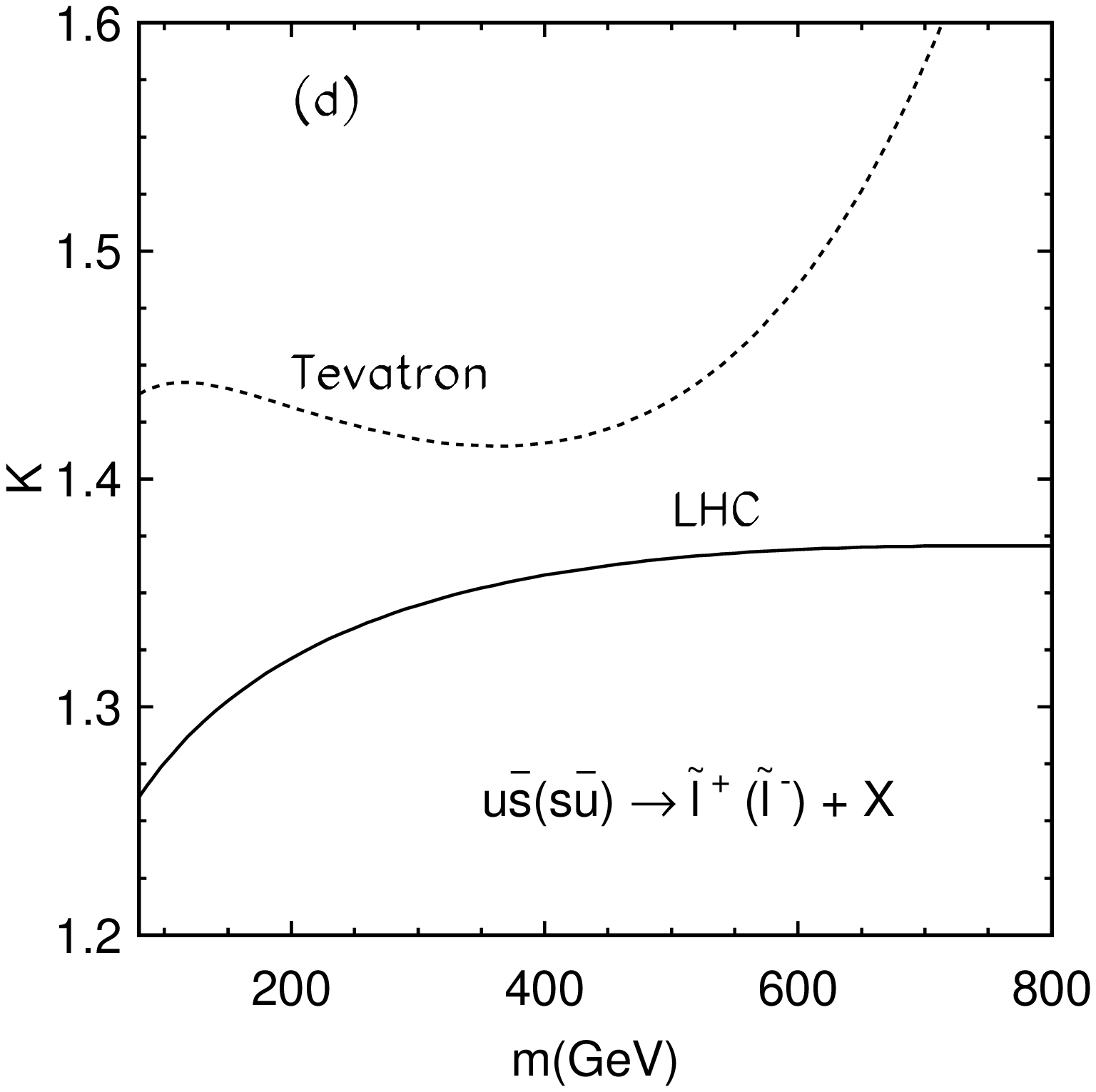,width=8cm}}
\caption{Cross sections at the LO (dashed line) and the
NLO (solid line) and  their corresponding  $K$-factor
related to ${\lambda'}_{i12}$  for $d\bar{s}(s\bar{d})\rightarrow \tilde{\nu}_i+X$
shown  in (a) and (c), and $u\bar{s}(s\bar{u})\rightarrow
\tilde{\ell}^+(\tilde{\ell}^-)$ in (b) and (d) at the Tevatron and the LHC.}
\label{csnlol12}
\end{figure}

\begin{figure}[tb]
\centerline{\psfig{figure=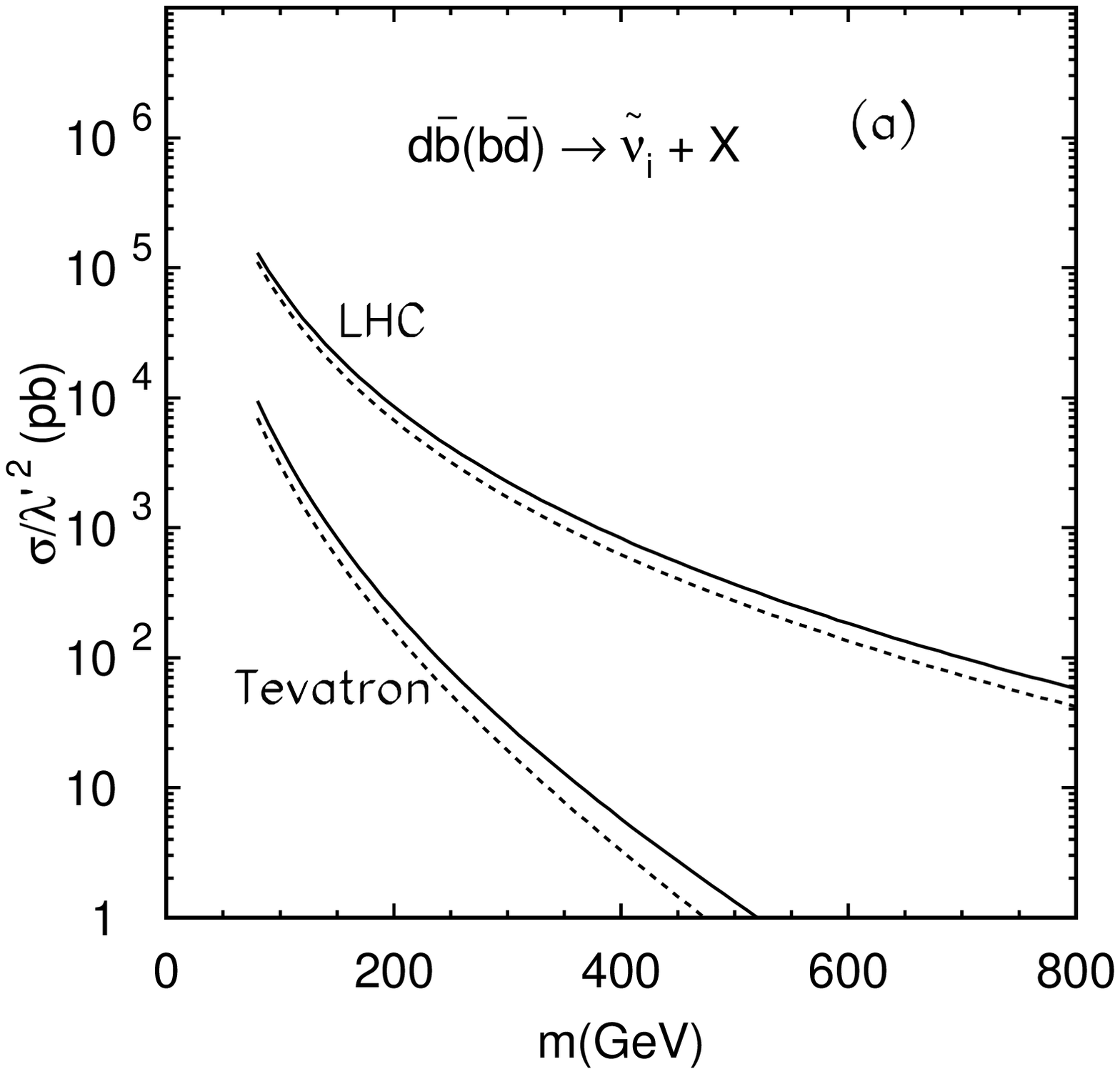,width=8cm}
\psfig{figure=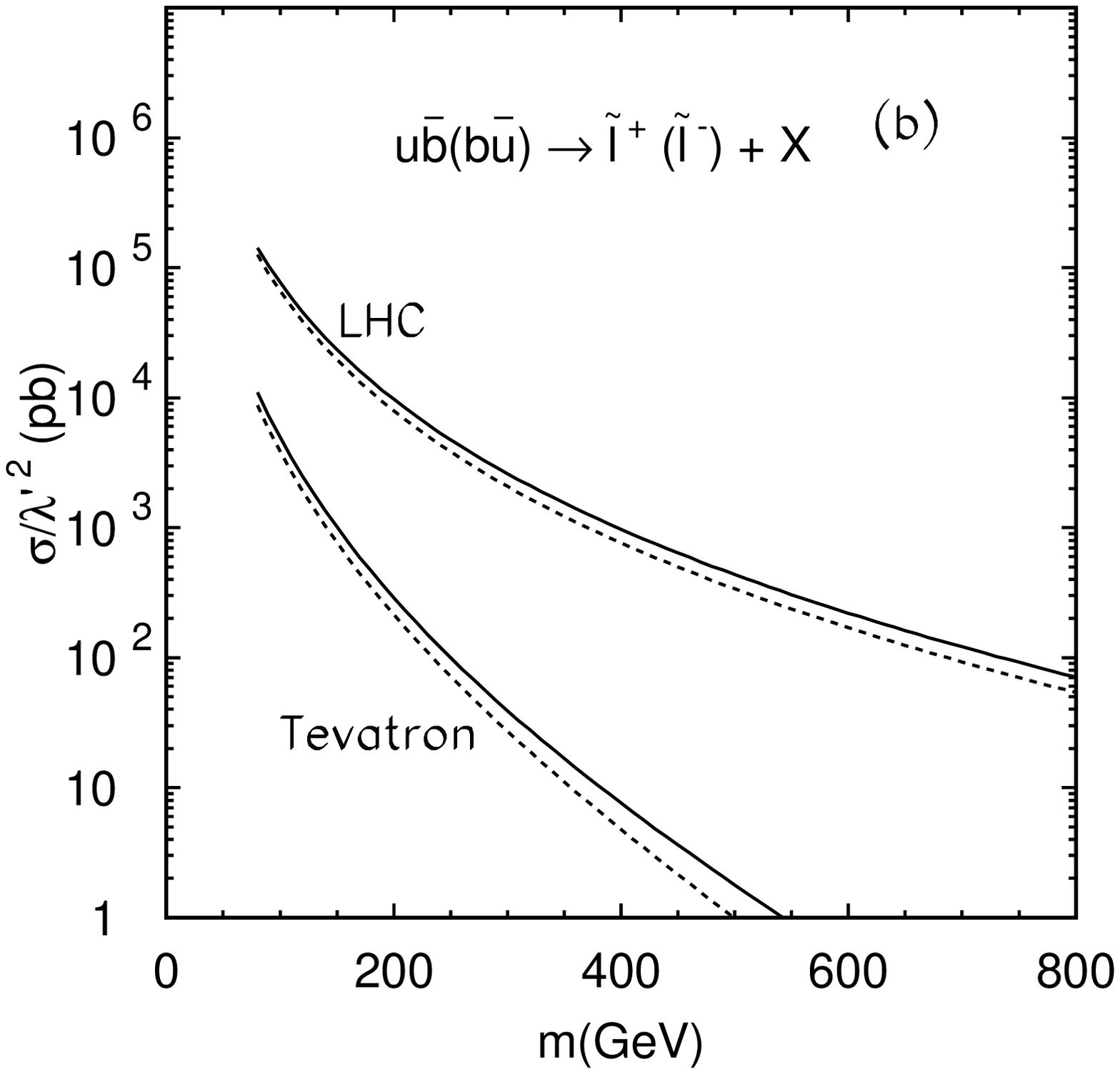,width=8cm}}
\centerline{\psfig{figure=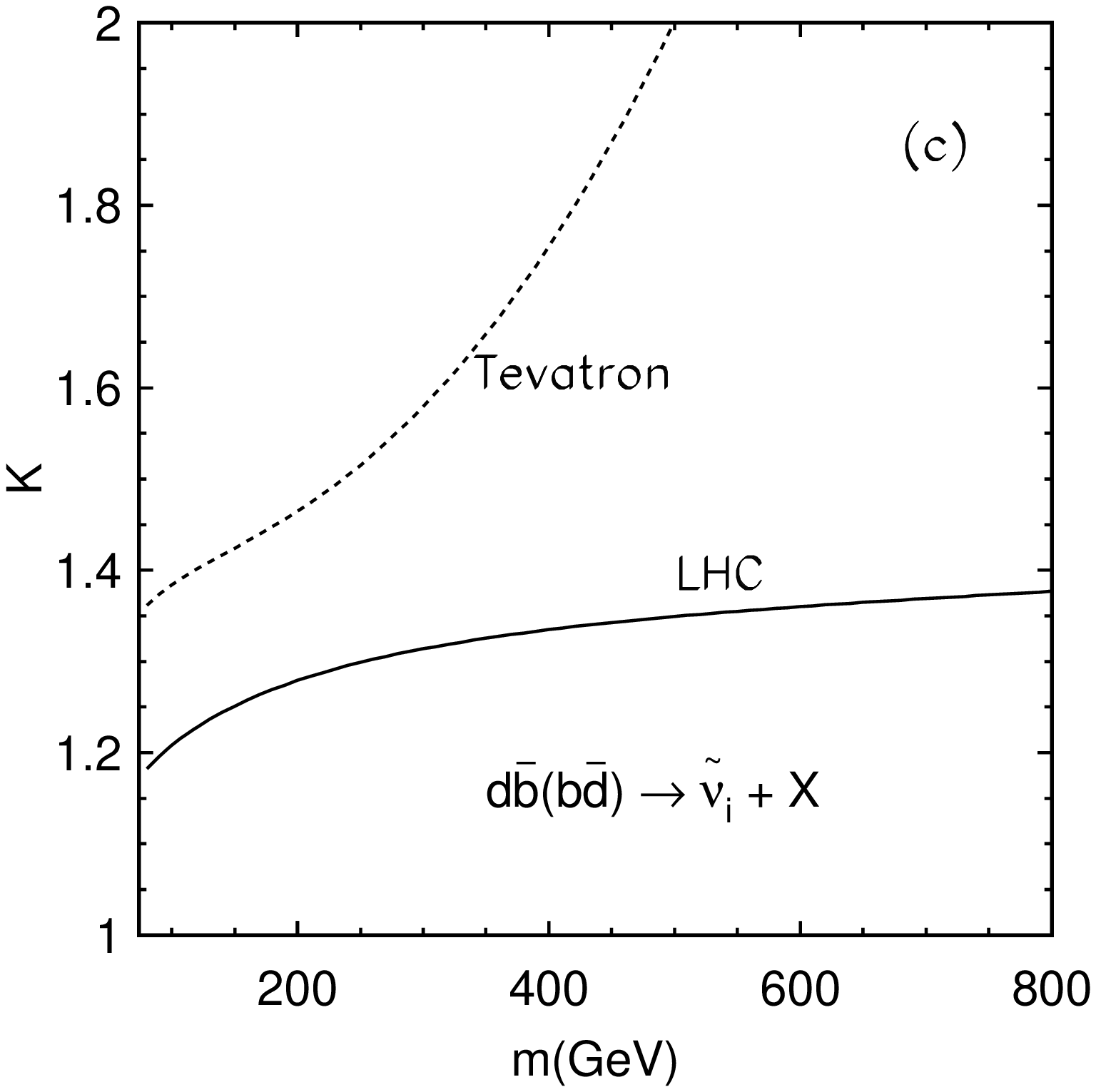,width=8cm}
\psfig{figure=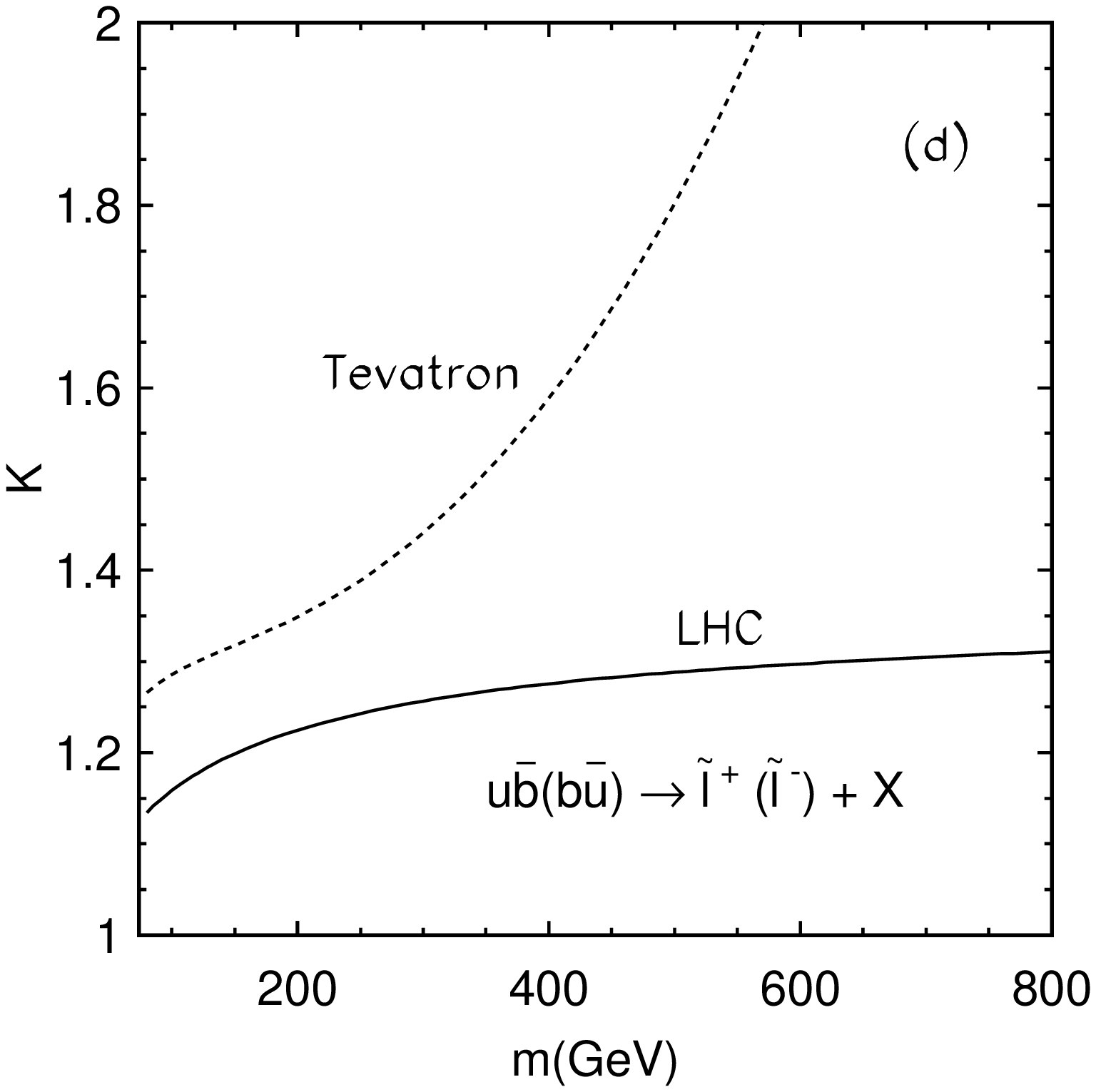,width=8cm}}
\caption{Cross sections at the LO (dashed line) and the NLO (solid line)
and  their corresponding  $K$-factor  related to ${\lambda'}_{i13}$ for
$d\bar{b}(b\bar{d})\rightarrow \tilde{\nu}_i+X$ shown in  (a) and (c), 
and $u\bar{b}(b\bar{u})\rightarrow
\tilde{\ell}^+(\tilde{\ell}^-)$ in (b) and (d) at the Tevatron 
and the  LHC.}
\label{csnlol13}
\end{figure}

\begin{figure}[tb]
\centerline{\psfig{figure=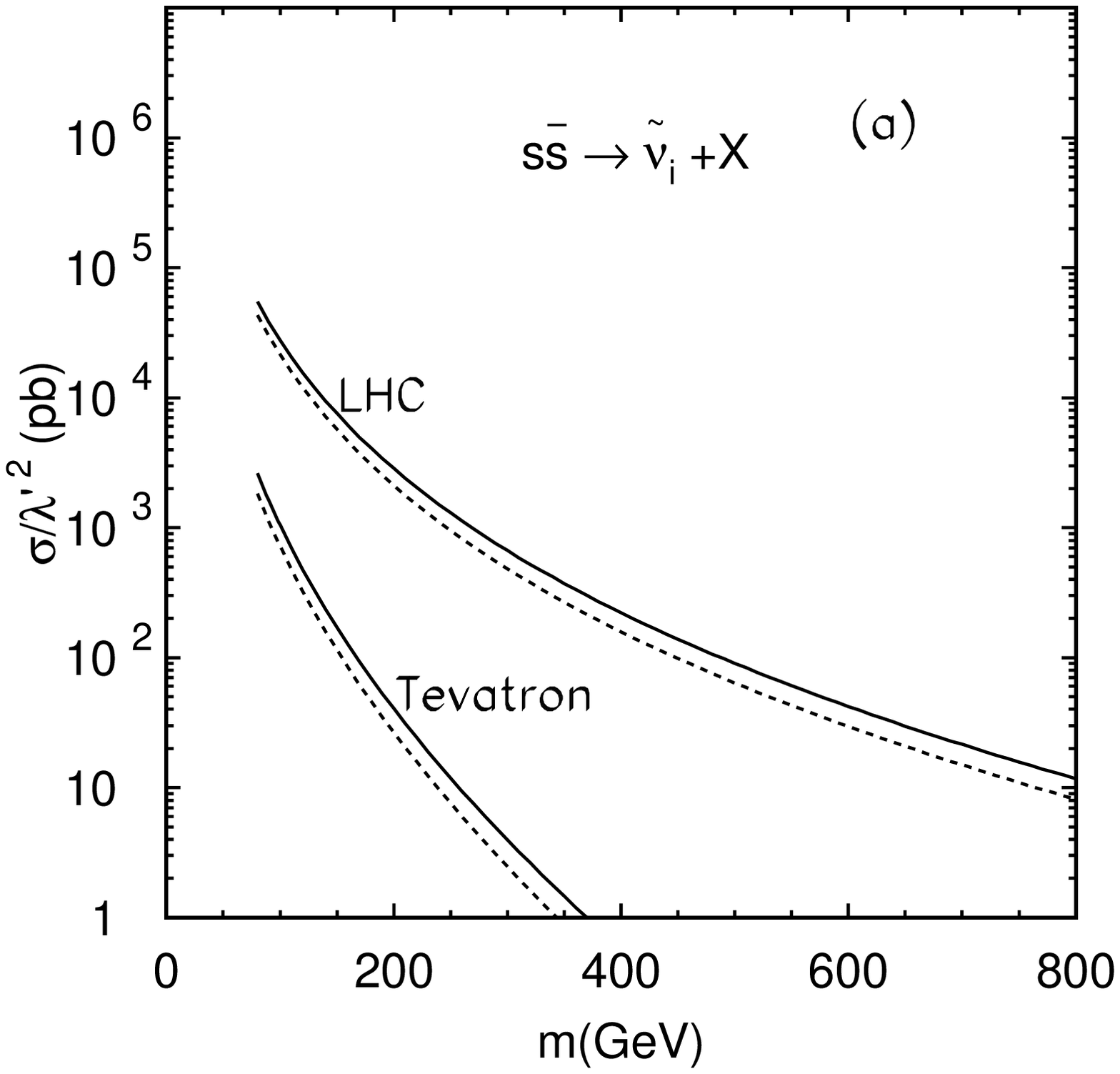,width=8cm}
\psfig{figure=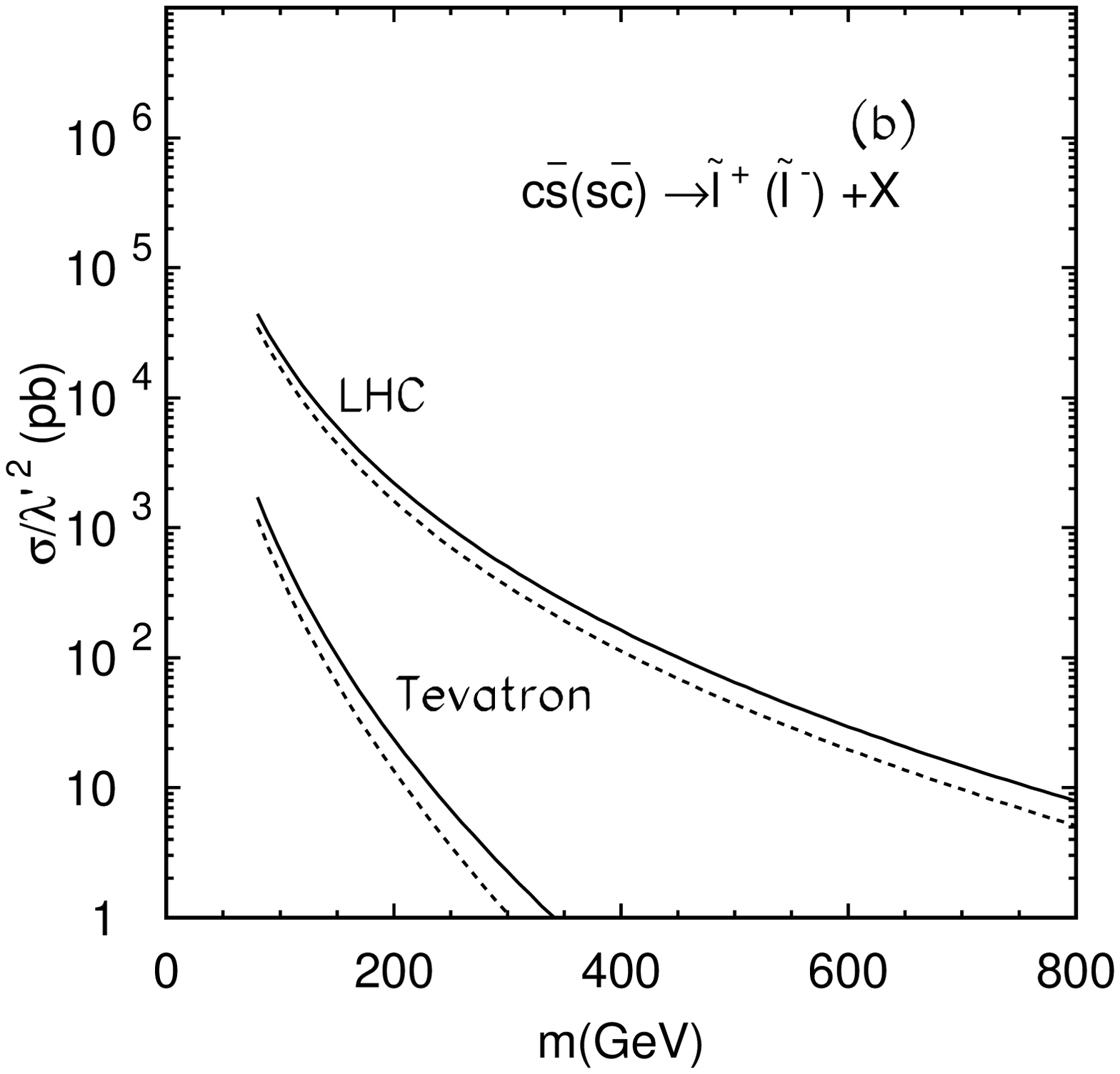,width=8cm}}
\centerline{\psfig{figure=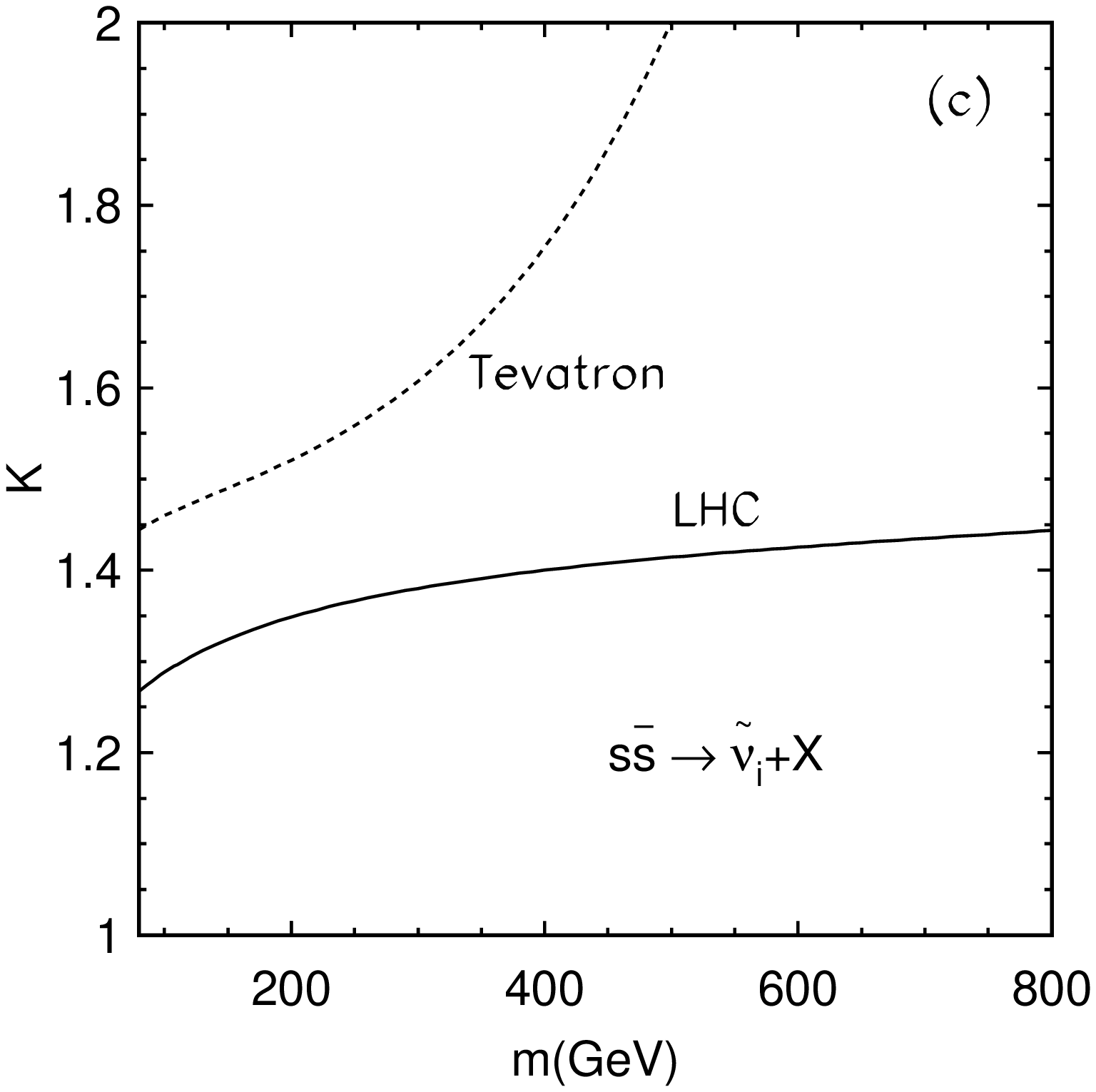,width=8cm}
\psfig{figure=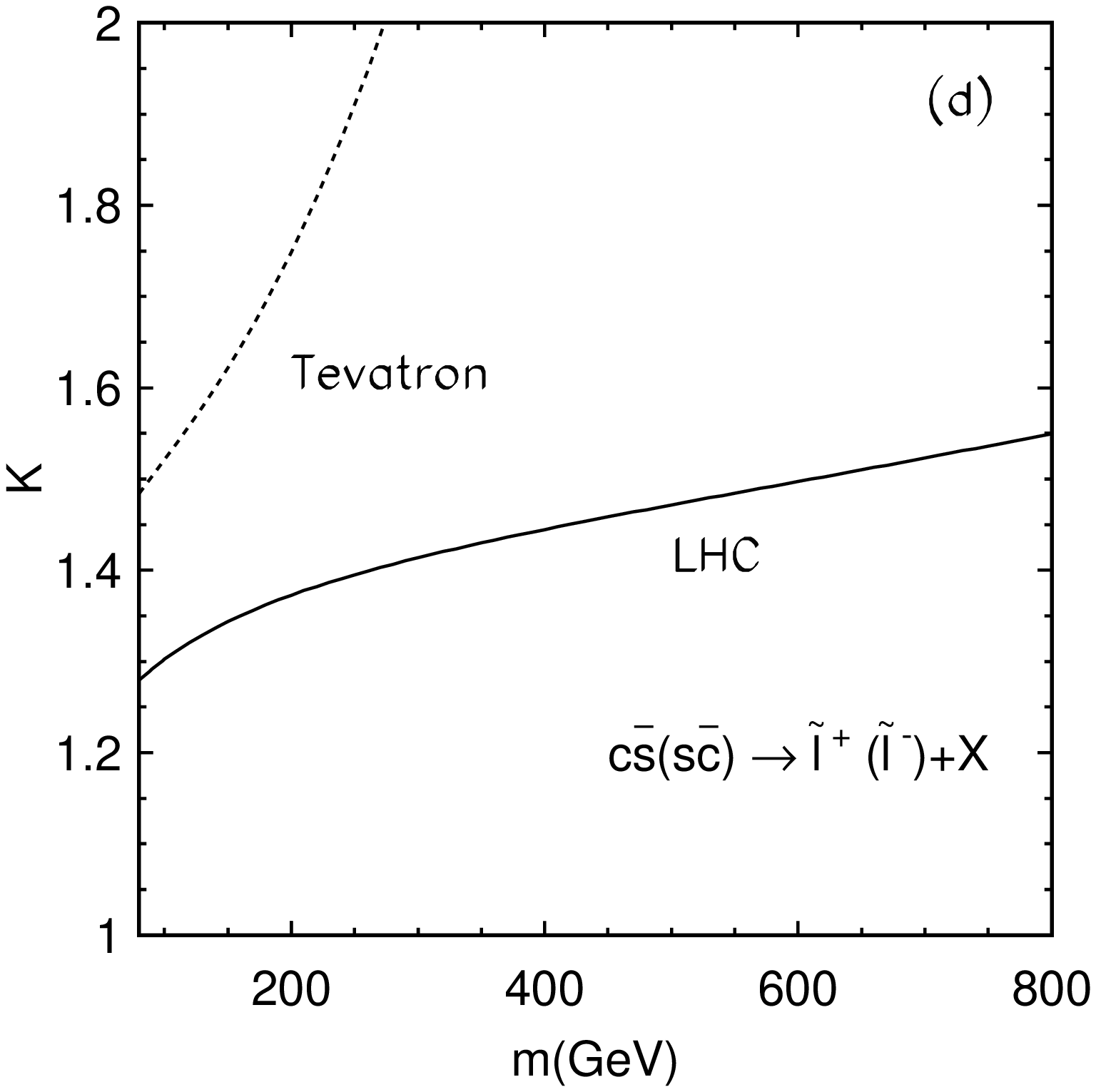,width=8cm}}
\caption{Cross sections at the LO (dashed line) and the NLO (solid line)
and  their corresponding  $K$-factor  related to ${\lambda'}_{i22}$ for
$s\bar{s} \rightarrow \tilde{\nu}_i+X$ shown in  (a) and (c), 
and $c\bar{s} (s\bar{c})\rightarrow
\tilde{\ell}^+(\tilde{\ell}^-)$ in (b) and (d) at the Tevatron 
and the  LHC.}
\label{csnlol22}
\end{figure}

\begin{figure}[tb]
\centerline{\psfig{figure=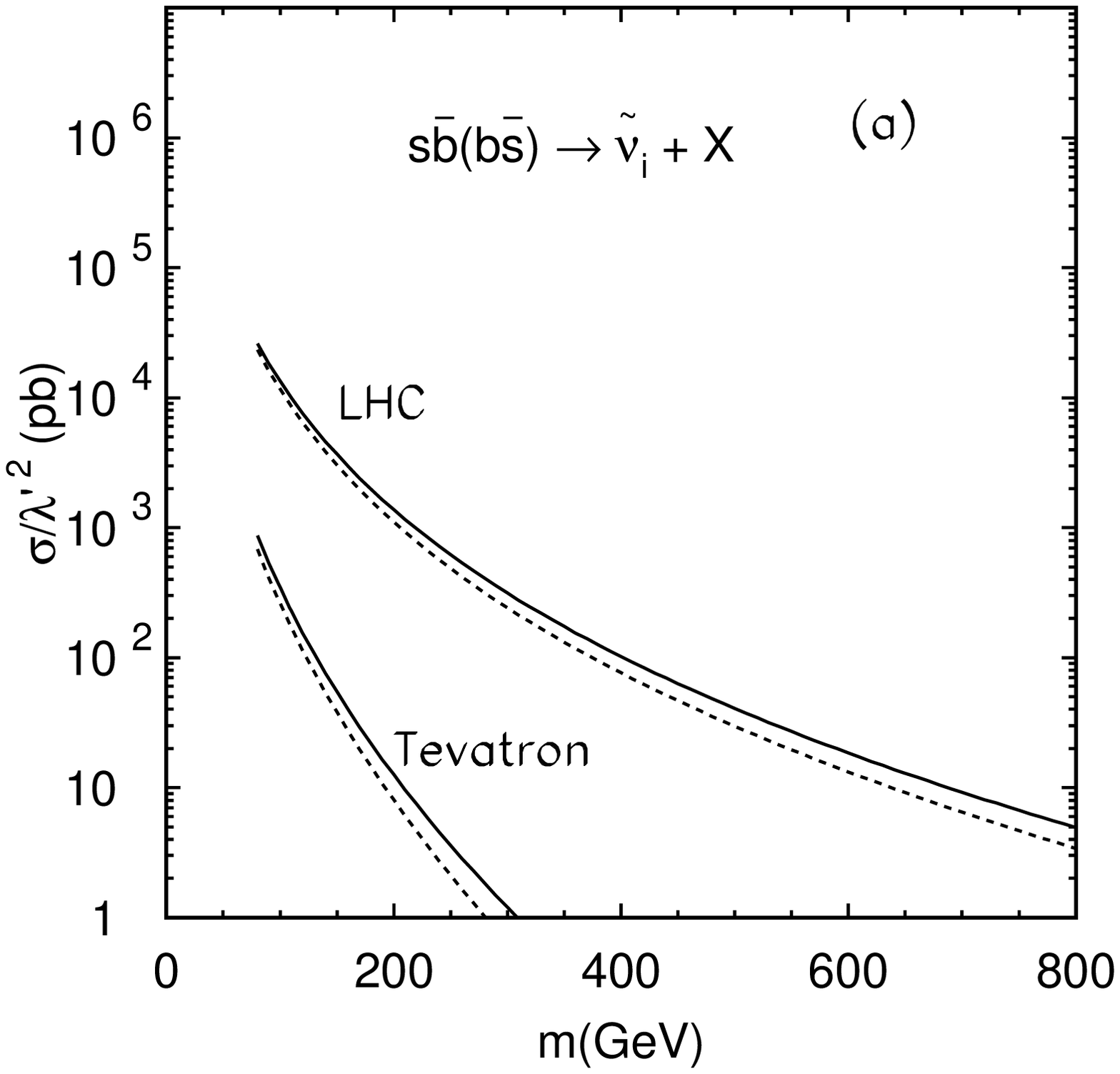,width=8cm}
\psfig{figure=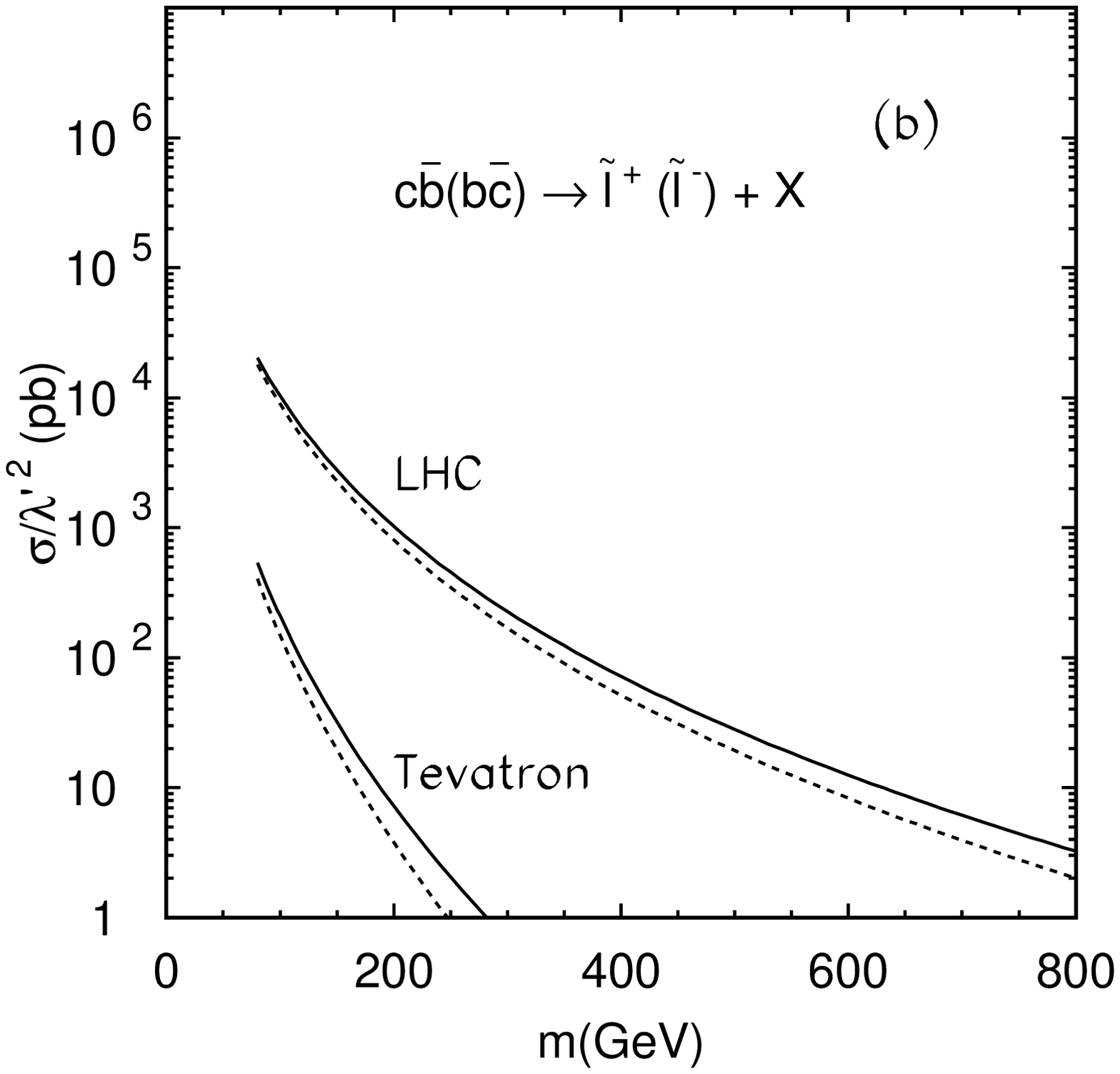,width=8cm}}
\centerline{\psfig{figure=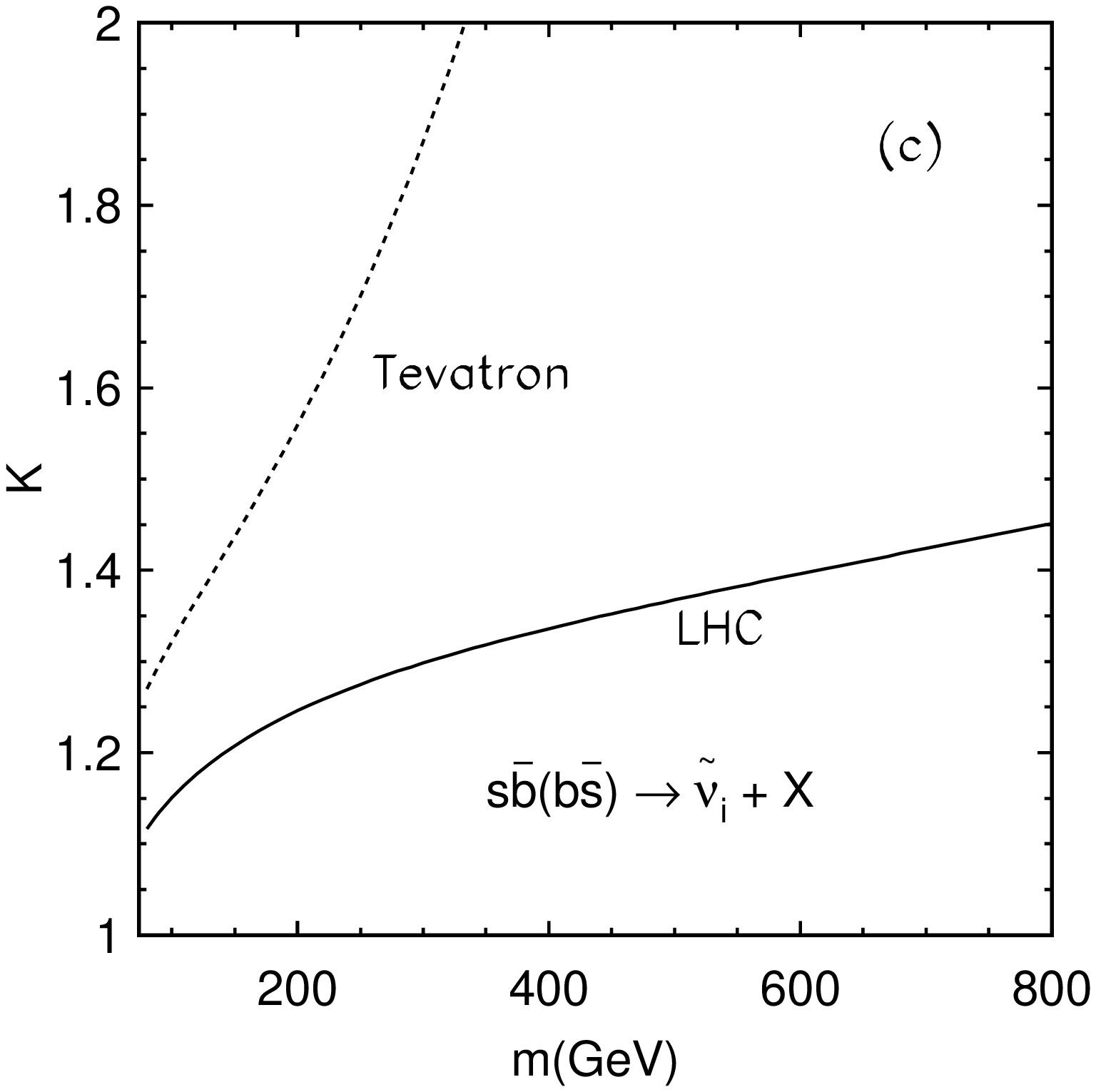,width=8cm}
\psfig{figure=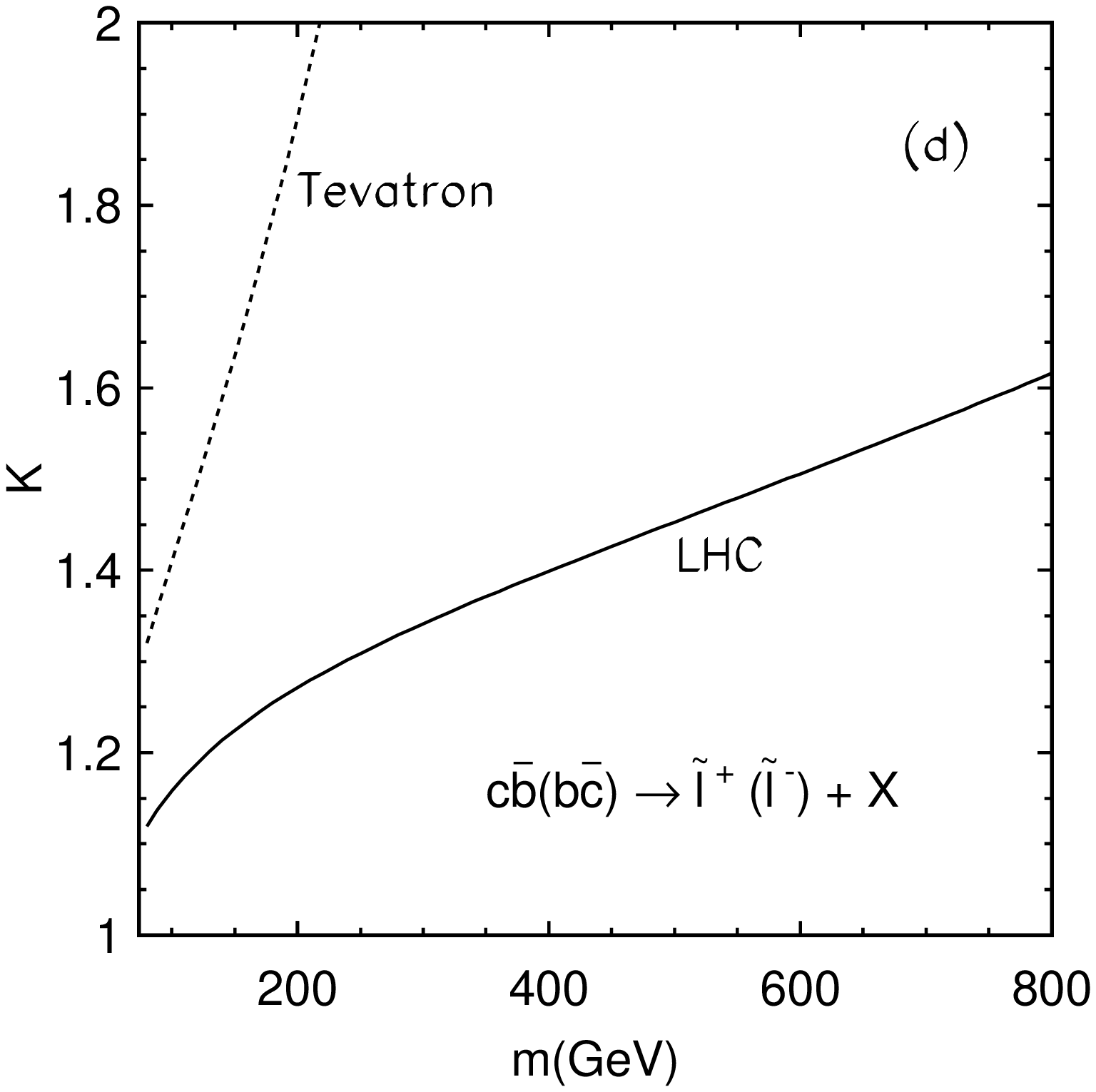,width=8cm}}
\caption{Cross sections at the LO (dashed line) and the NLO (solid line)
and  their corresponding  $K$-factor  related to ${\lambda'}_{i23}$ for
$s\bar{b}(b\bar{s})\rightarrow \tilde{\nu}_i+X$ shown in  (a) and (c), 
and $c\bar{b}(b\bar{c})\rightarrow
\tilde{\ell}^+(\tilde{\ell}^-)$ in (b) and (d) at the Tevatron 
and the  LHC.}
\label{csnlol23}
\end{figure}

\begin{figure}[tb]
\centerline{\psfig{figure=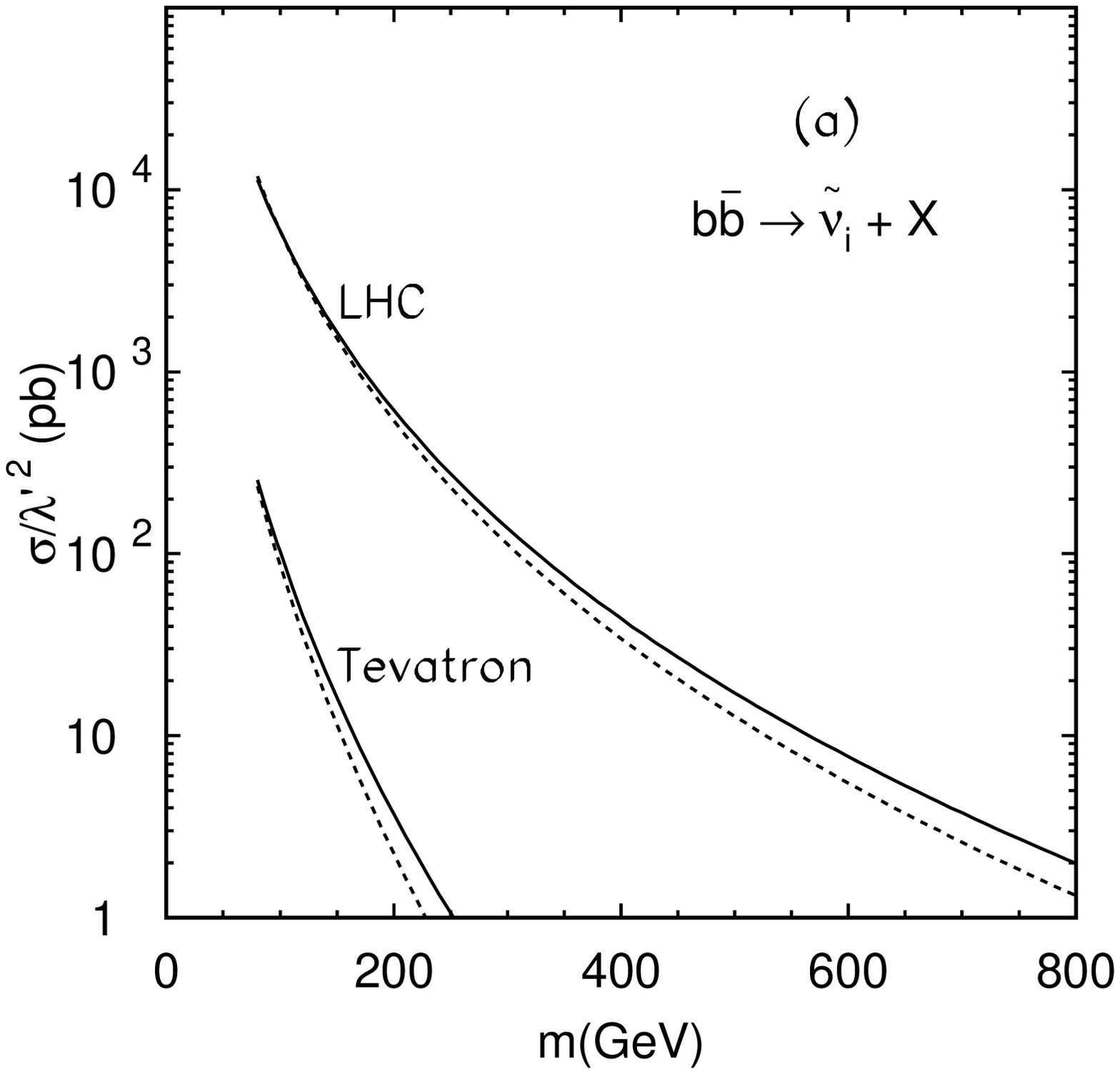,width=8cm}
\psfig{figure=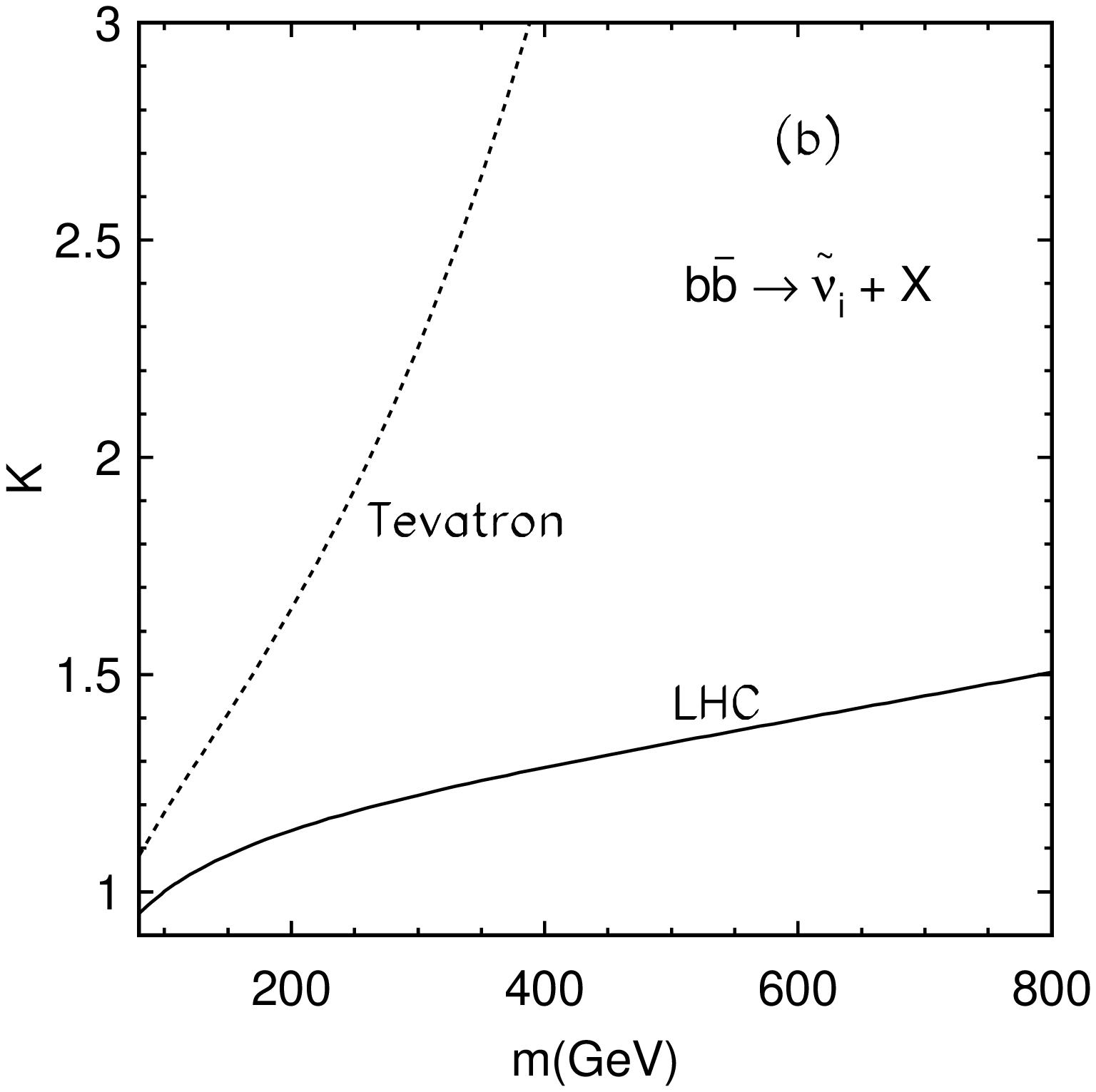,width=8cm}}
\caption{(a) Cross sections related to ${\lambda'}_{i33}$ 
for $b\bar{b}\rightarrow \tilde{\nu}_i+X$ at the
LO (dashed line) and the NLO (solid line) at the Tevatron and the LHC
and (b) their corresponding $K$-factor.}
\label{csnlol33}
\end{figure}

To keep our presentation  model-independent,
we pull out the overall  $\rpv$ coupling  ${\lambda'}_{ijk}^2$
when showing the cross sections unless specified otherwise.
We remind the reader that the last two indices ($j,k$) refer to the
generation of quark partons and the first index ($i$) is for the slepton.
Thus our results are formally applicable to any flavor of the sleptons,
and are given in Figs.~\ref{csnlol11}$-$\ref{csnlol33}:
\begin{eqnarray}
&& {\rm Fig.~\ref{csnlol11}}: \quad
d\bar{d} \rightarrow \tilde{\nu}_i,\ \  {\rm and}\ \
u\bar{d}\ (d\bar{u})\rightarrow \tilde{\ell}^+\ (\tilde{\ell}^-),\quad
{\rm for\ probing}\ \lambda'_{i11}.
\nonumber \\
&& {\rm Fig.~\ref{csnlol12}}: \quad
d\bar{s},\ s\bar{d} \rightarrow \tilde{\nu}_i,\ \  {\rm and}\ \
u\bar{s}\ (s\bar{u})\rightarrow \tilde{\ell}^+\ (\tilde{\ell}^-),\quad
{\rm for\ probing}\ \lambda'_{i12}.
\nonumber \\
&& {\rm Fig.~\ref{csnlol13}}: \quad
d\bar{b},\ b\bar{d} \rightarrow \tilde{\nu}_i,\ \  {\rm and}\ \
u\bar{b}\ (b\bar{u})\rightarrow \tilde{\ell}^+\ (\tilde{\ell}^-),\quad
{\rm for\ probing}\ \lambda'_{i13}.
\nonumber \\
&& {\rm Fig.~\ref{csnlol22}}: \quad
s\bar{s} \rightarrow \tilde{\nu}_i,\ \  {\rm and}\ \ 
c\bar{s}\ (s\bar{c})\rightarrow \tilde{\ell}^+\ (\tilde{\ell}^-),\quad
{\rm for\ probing}\ \lambda'_{i22}.
\nonumber \\
&& {\rm Fig.~\ref{csnlol23}}: \quad
s\bar{b},\ b\bar{s} \rightarrow \tilde{\nu}_i,\ \  {\rm and}\ \
c\bar{b}\ (b\bar{c})\rightarrow \tilde{\ell}^+\ (\tilde{\ell}^-),\quad
{\rm for\ probing}\ \lambda'_{i23}.
\nonumber \\
&& {\rm Fig.~\ref{csnlol33}}: \quad
b\bar{b} \rightarrow \tilde{\nu}_i, \quad
{\rm for\ probing}\ \lambda'_{i33}.
\nonumber
\end{eqnarray}
It is seen that the production rate of
the slepton is larger than that of sneutrino, due to the parton luminosity difference
and the charged state counting. 
While the production cross sections for these processes at the LHC 
energies are larger than that at the Tevatron energies
by about one order of magnitude at $m_{\tilde\ell} \lsim 200$ GeV, 
and about two orders of magnitude at $m_{\tilde\ell} \gsim 700$ GeV.
we see that the NLO QCD corrections and the $K$-factors are 
largely reflecting ratio of the parton distributions at different 
scale and $x$-values, and are of the following features.

\begin{itemize}
\item At the LHC, the QCD corrections are rather stable and 
increasing monotonically versus $m_{\tilde\ell}$. 
The $K$-factors are typically around $1.2-1.4$, except that the
range becomes slightly larger when more sea quarks are involved.
\item At the Tevatron, the $K$-factor is around $1.1-1.5$ when 
the valence quarks dominate as for $\lambda_{i11}$ in Fig.~\ref{csnlol11}. 
For other channels, the QCD corrections are less stable.
The $K$-factors  can be larger than a factor of two at high masses
or large $x$-values where the sea quark distributions fall
quickly. This effect was also observed in \cite{Choudhury2003}.
\end{itemize}

%

\subsection{Event rates and signal detection at hadron colliders}

So far we  have not  specified the values for the  $\rpv$ couplings.
There are stringent  bounds on them from low energy data. 
We consider the strongest constraints for $\lambda'_{ijk}$
determined from the charge and neutral currents as summarized 
in a recent review \cite{Barbier04ez},
and calculate the corresponding cross sections
for the slepton and sneutrino production. 
The results are listed in Table \ref{tabslep}. It is known that the bounds
on the couplings of the lighter generation sleptons are significantly 
tighter than the heavier generations.
The production rates related to $\tilde{\tau}$
and $\tilde{\nu}_{\tau}$ could be much larger than that of the
other sleptons and sneutrinos because of the possibly larger couplings.
For illustration, 
we have considered the mass range of $m_{\tilde\ell}=150-500$ GeV.
The production rates are sizable, given the fact that the designed high luminosity at
the LHC may reach 100 fb$^{-1}$ annually.

As for the final state identification of the signal, it is likely to go through the
$R$-parity conserving decays as the  $\lambda's$ are expected to be small. 
Thus the clean decay modes with at least one charged lepton may include
\begin{eqnarray}
\tilde\ell^\pm & \to & \tilde\chi^0 \ell^\pm,\ \tilde\chi^\pm \nu; \\
\tilde\nu & \to & \tilde\chi^\pm \ell^\pm .
\end{eqnarray}
The feasibility for the signal identification has been  studied
extensively in the literature based on leading order cross sections,
and we thus refer the readers to a review article  \cite{Barbier04ez} 
and references therein.
Especially,  the appropriate coupling ranges for neutrino mass generation 
can be tested in the collider experiments \cite{Rnew}.  It is also  interesting
to note that for a sizable coupling $\lambda'_{i33}$, it may lead to 
an observable heavy quark final state $b\bar b$, but not $t\bar t$.

\begin{table}[tb]
\begin{center}
\begin{tabular}{|l|c|cll|} \hline
$\lambda'$ & Upper    &\multicolumn{3}{c|}{ $\sigma$ for $m_{\tilde\ell}=150-500$\ GeV } \\
                     & bounds  & &  Tevatron (pb)      & LHC (pb)  \\ \hline
$\lambda'_{111} $  & 0.02    &$\tilde{\nu}_e$  & 1.0--0.01 &
10--0.21   \\\cline{3-5}
                   &         &$\tilde{e}$      & 3.6--0.03 & 25--0.55   \\ \hline
$\lambda'_{211} $  & 0.06    &$\tilde{\nu}_{\mu}$  & 9.8--0.05 &
91--1.9 \\ \cline{3-5}
                   &         &$\tilde{\mu}$        &33--0.28 & 221--5.0 \\ \hline
$\lambda'_{311} $  & 0.12    &$\tilde{\nu}_{\tau}$  & 36--0.21 &
363--7.6 \\  \cline{3-5}
                   &         &$\tilde{\tau}$        &131--1.1 & 885--20 \\ \hline
$\lambda'_{112} $  & 0.02    &$\tilde{\nu}_e$  & 0.59--$10^{-3}$ &
12--0.21   \\ \cline{3-5}
                   &         &$\tilde{e}$      &1.37--3$\cdot 10^{-3}$    & 22--0.42   \\ \hline
$\lambda'_{212} $  & 0.06    &$\tilde{\nu}_{\mu}$  & 5.3--0.01 &
110--1.9  \\ \cline{3-5}
                   &         &$\tilde{\mu}$        &12--0.03 & 201--3.7 \\ \hline
$\lambda'_{312} $  & 0.12    &$\tilde{\nu}_{\tau}$  & 21--0.03 &
442--7.5   \\ \cline{3-5}
                   &         &$\tilde{\tau}$   & 49--0.10      &805--15 \\ \hline
$\lambda'_{113} $  & 0.02    &$\tilde{\nu}_e$  &0.33--5$\cdot
10^{-4}$ & 8.4--0.15 \\ \cline{3-5}
                   &         &$\tilde{e}$      &0.40--7$\cdot 10^{-4}$  & 9.4--0.17 \\ \hline
$\lambda'_{213} $  & 0.06    &$\tilde{\nu}_{\mu}$  &3.0--5 $\cdot
10^{-3}$ & 76--1.3 \\ \cline{3-5}
                   &         &$\tilde{\mu}$        &3.6--0.01  &85--1.6 \\ \hline
$\lambda'_{313} $  & 0.12    &$\tilde{\nu}_{\tau}$ &12--0.02 &302--5.8 \\
                   &         &$\tilde{\tau}$     &14--0.03  & 339--6.3 \\  \hline
$\lambda'_{122} $  & 0.21    &$\tilde{\nu}_e$  & 7.5--4$\cdot
10^{-3}$ & 333--4.0   \\ \cline{3-5}
                   &         &$\tilde{e}$   & 9.1--5$\cdot 10^{-3}$   & 526--5.7  \\ \hline
$\lambda'_{222} $  & 0.21    &$\tilde{\nu}_{\mu}$  & 7.5--4$\cdot
10^{-3}$ & 333--4.0   \\ \cline{3-5}
                   &         &$\tilde{\mu}$   & 9.1--5$\cdot 10^{-3}$   & 526--6.0   \\ \hline
$\lambda'_{322} $  & 0.52    &$\tilde{\nu}_{\tau}$  & 46--0.03
& 2043--24 \\ \cline{3-5}
                   &         &$\tilde{\tau}$        & 56--0.03  & 3223-35  \\ \hline
$\lambda'_{123} $  & 0.21    &$\tilde{\nu}_e$  & 2.4--1$\cdot
10^{-3}$ & 162--1.8
\\ \cline{3-5}
                   &         &$\tilde{e}$      & 2.8--2$\cdot 10^{-3}$  &  245--2.5 \\ \hline
$\lambda'_{223} $  & 0.21    &$\tilde{\nu}_{\mu}$  & 2.4--1$\cdot
10^{-3}$ & 162--1.8   \\ \cline{3-5}
                   &         &$\tilde{\mu}$      & 2.8--2$\cdot 10^{-3}$  &  245--2.5 \\ \hline
$\lambda'_{323} $  & 0.52    &$\tilde{\nu}_{\tau}$  &  15--0.01 &
992--11 \\  \cline{3-5}
                   &         &$\tilde{\tau}$        &  17--0.01  & 1502--15 \\ \hline
$\lambda'_{133} $  & 0.18    &$\tilde{\nu}_e$  &  0.52--3$\cdot 10^{-4}$   &
53--0.56   \\ \hline
 $\lambda'_{233} $  & 0.45 &$\tilde{\nu}_{\mu}$
& 3.3--2$\cdot 10^{-3}$ & 333--3.5   \\ \hline
$\lambda'_{333} $ &
0.58    &$\tilde{\nu}_{\tau}$ & 5.4--3$\cdot 10^{-3}$ & 553--5.8   \\
\hline
\end{tabular}
\end{center}
\caption{ Upper bounds on $\lambda'$ in units of $(m_{\tilde\ell}/100\ \gev)^{-2}$
and the corresponding cross sections for slepton and sneutrino production.}
\label{tabslep}
\end{table}


\section{Transverse momentum distributions and soft gluon resummation}
\label{sec4}

With the differential cross sections obtained in Sec.~3, we can also calculate
the transverse momentum ($p_T$) distribution of the produced slepton
associated with a hard parton in the final state. 
In considering the $p_T$ distribution,
two different energy scales are involved, i.e., $p_T$ and $m_{\tilde\ell}$.
 When $p_T$ is comparable to $m_{\tilde\ell}$ or
higher, fixed order perturbation theory gives sufficiently
accurate results. However, when $p_T \ll m_{\tilde\ell}$,
large logarithmic terms
such as $(\alpha_s \ln^2 p_T/m_{\tilde\ell})^n$ arise at fixed order
perturbation calculations and need to be resummed in order to obtain reliable
predictions.

Following the standard procedure \cite{Han:1991sa}, we first obtain
 the leading order asymptotic cross section
in the limit of small transverse momentum
\begin{eqnarray}
\frac{d\sigma^{asym}}{dp_T ~ dy}&=&\sigma_{Born}\,
\frac{\alpha_s}{\pi} \frac{1}{S\, p_T}\Big\{ \Big( A\ln\frac{m_{\tilde{\ell}}^2}{p_T^2}+B\Big)
f_q (x_1^0) f_{\bar{q}}(x_2^0)
+ (f_{\bar{q}} \otimes P_{\bar{q}\leftarrow \bar{q}})(x_1^0) f_q(x_2^0) \nonumber \\
&&+(f_g \otimes P_{\bar{q}\leftarrow g})(x_1^0) f_q(x_2^0)
\Big\}+\Big(x_1^0\leftrightarrow x_2^0\Big),\\
(f \otimes g)(x)& \equiv &\int\limits_x^1 f(y) \, g(\frac{x}{y}) dy, \ \
P_{\bar{q}\leftarrow \bar{q}}(x)=\frac{4}{3} \Big[ \frac{1+x^2}{1-x}\Big]_+, \ \
P_{\bar{q}\leftarrow g}(x) =\frac{1}{2} \Big[x^2+(1-x)^2\Big].
\nonumber
\end{eqnarray}
where $y$ is the rapidity of the parton c.m.~system,
$x^0_{1,2} \equiv e^{\pm y} m_{\tilde{\ell}}/\sqrt{S}$, and the expansion
coefficients relevant for the leading-log resummation are determined to be
$A=A^{(1)}=2C_F,\  B=B^{(1)}= -3 C_F$.

The resummed differential cross section can be expressed as the
integral over the impact parameter $b$:
\begin{eqnarray}
\frac{d\sigma^{resum}}{ d p_T dy}&=&\sigma_{Born} 
\frac{2P_T}{S}\int\limits_0^{\infty}  db \, \frac{b}{2} \, J_0(b p_T) W(b),\\
W(b)&=&\exp\left[\, 
-\int\limits_{{b_0^2}/{b_*^2}}^{m_{\tilde{l}}^2}\,\frac{dq^2}{q^2}
\, \frac{\alpha_s(q^2)}{2\pi} \, \Big(
A\ln\frac{m_{\tilde{l}}^2}{p_T^2}+B\Big)\,\right]\,
S_{np} f_q(x_1^0)f_{\bar{q}}(x_2^0)+\Big(x_1^0\leftrightarrow x_2^0\Big), \nonumber\\
S_{np}&=&\exp\left(-b^2 g_1-b^2 g_2\, \ln \,\frac{b_{max}
m_{\tilde{l}}}{2} \, \right)\,,
\nonumber\\
b_*&=&\frac{b}{1+b^2/b_{max}^2}, ~~g_1=0.24\ {\gev}^2, ~~g_2=0.34\
{\gev}^2,~~b_{max}=(1\ {\gev})^{-1},
\end{eqnarray}
where the parameters are taken from Ref.~\cite{cpy} based on a fit to the
Tevatron Drell-Yan data.
Including an ansaz for  matching between the resummed low momentum region
and the perturbative high  momentum region,
the full transverse momentum distribution is thus expressed as 
\begin{eqnarray}
\frac{d\sigma^{full}}{dp_T}&=&\int dy \, \left[ \,
\frac{d\sigma^{pert}}{dp_T dy} +f(p_T)\,\left( \,
\frac{d\sigma^{resum}}{d p_T dy} -\frac{d\sigma^{asym}}{dp_T ~dy}\, \right)\, \right],\\
f(p_T)&=&\frac{1}{1+(p_T/m_{\tilde{l}})^4}.
\end{eqnarray}
We must emphasize that the ad-hoc function $f(p_T^{})$ is somewhat arbitrary and
it reflects our ignorance for the distribution in the region $p_T\sim m_{\tilde{l}}$.
The only way to reduce this arbitrariness would be to include higher order
calculations in both perturbative and resummed regimes. 

\begin{figure}[tb]
\centerline{
\psfig{figure=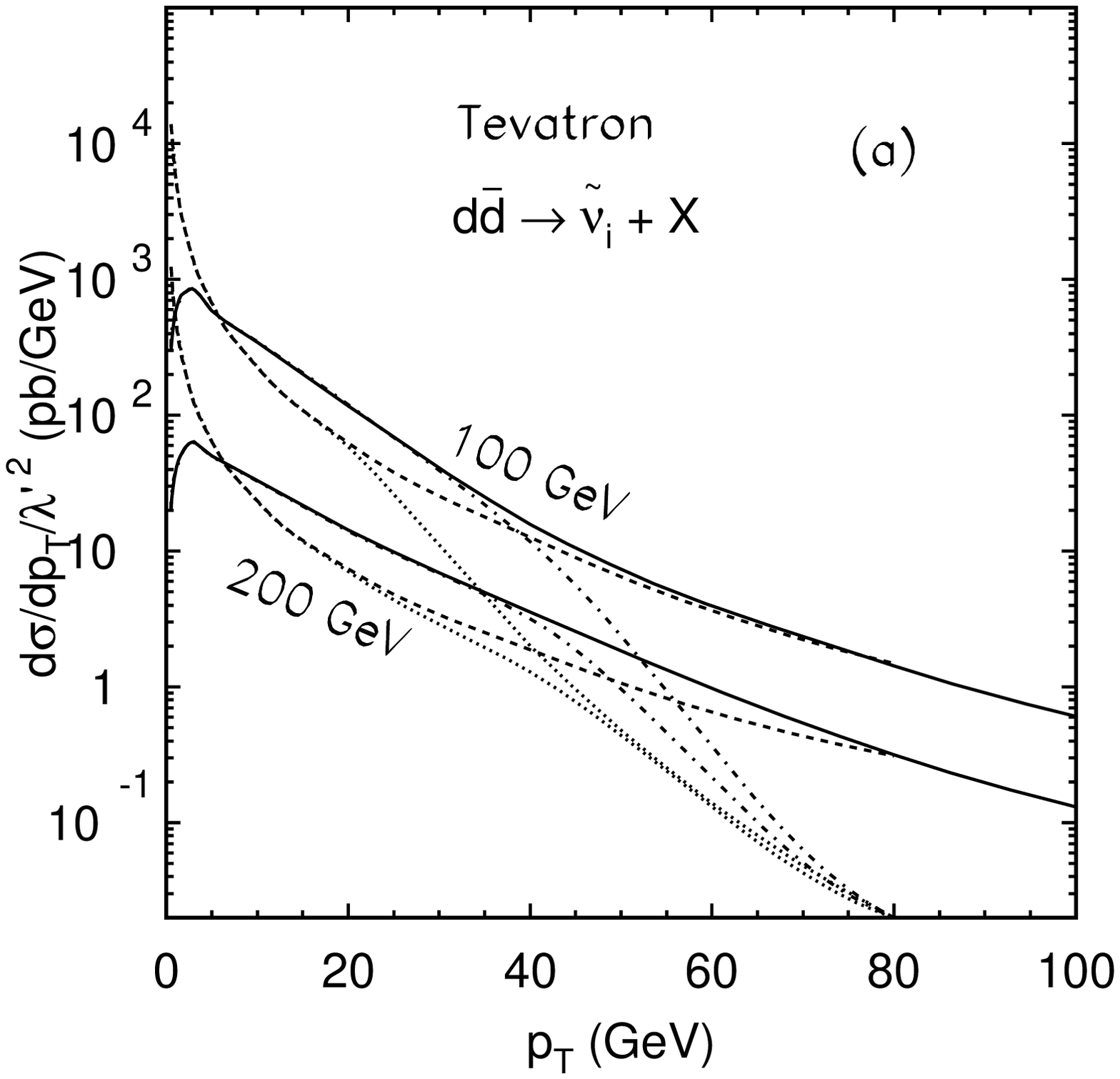,width=8cm}
\psfig{figure=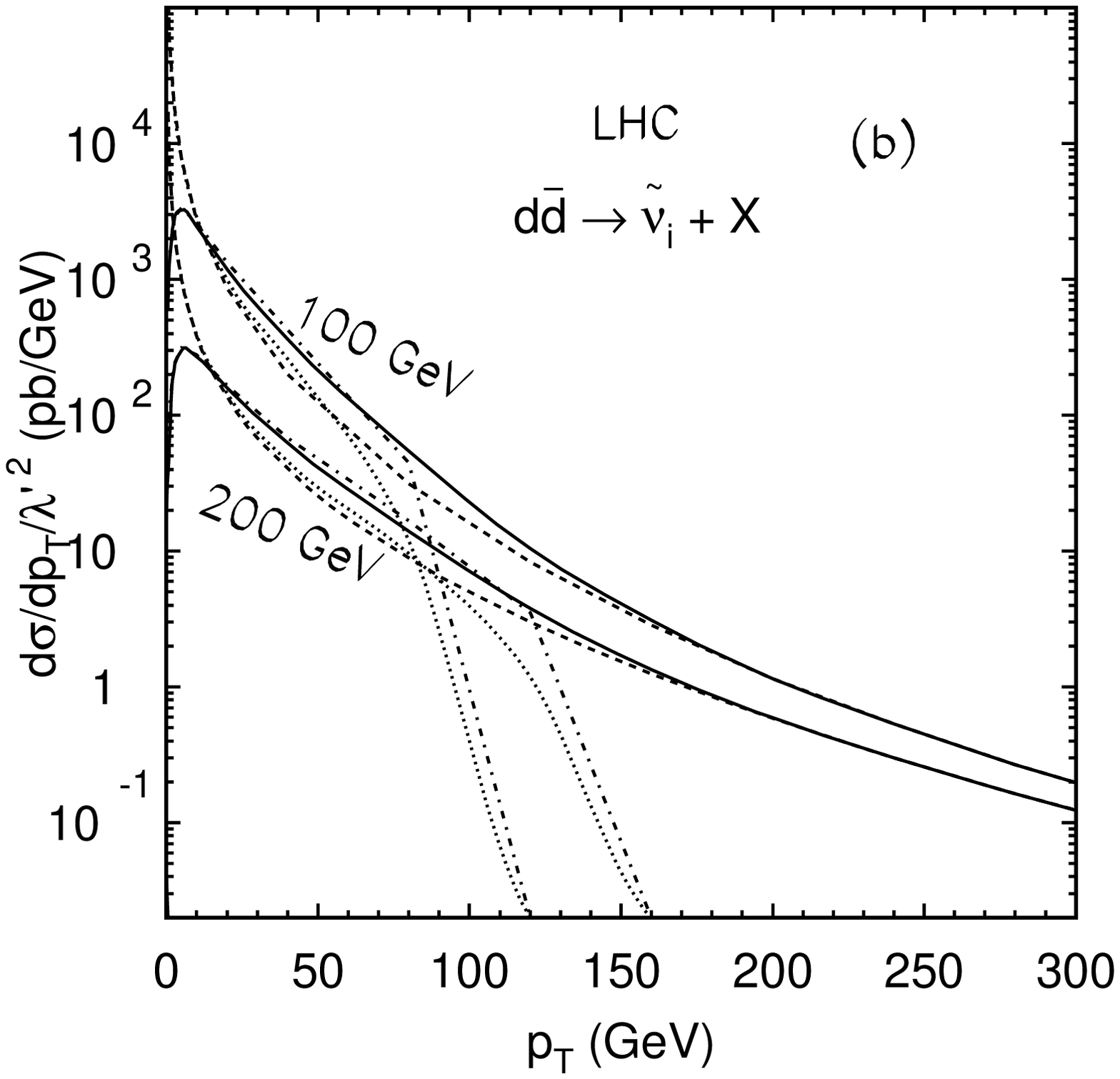,width=8cm}}
 \caption{Distributions of ${d\sigma^{full}}/{dp_T}$ (solid line),
${d\sigma^{pert}}/{dp_T}$ (dashed line), ${d\sigma^{resum}}/{dp_T}$
(dash-dotted line) and ${d\sigma^{asym}}/{dp_T}$ (dotted line) for
sneutrino production at (a) Tevatron ($\sqrt{s}=2$ TeV), 
and (b) LHC ($\sqrt{s}=14$ TeV).} \label{dpt1}
\end{figure}

Here as an example, we show the results of $d\bar{d}\rightarrow \tilde{\nu}_i+X$
in Fig.~\ref{dpt1}, with $m_{\tilde{\nu}_i}=100$ and $200$ GeV at
Tevatron ($\sqrt{s}=2$ TeV) and LHC ($\sqrt{s}=14$ TeV).
The results are obtained by using CTEQ6.1M. The distribution at
small $p_T$ is dominated by the resummed part, while at large
$p_T$, the contributions are mostly from the perturbative part.
For the $\frac{d\sigma^{full}}{dp_T}$, the peak is between $2$ and $3$ GeV.
The results for the other $q\bar{q}'\rightarrow \tilde{\ell}+X$ processes are similar.

\section{The $gg \rightarrow \tilde{\nu}$ Process}
\label{sec5}

Because of the large gluon luminosity at high energies, one may consider
the gluon fusion contribution to the single slepton production via quark triangle
loops
\begin{equation}
g(p_1)+g(p_2)\rightarrow \tilde{\nu}(q).
\end{equation}

\begin{figure}[tb]
\begin{center}
\begin{picture}(200,100)(0,0)
\SetWidth{1.8} \Gluon(20,20)(60,20){3}{5}
\Gluon(20,80)(60,80){3}{5} \ArrowLine(60,20)(60,80)
\ArrowLine(60,80)(100,50) \ArrowLine(100,50)(60,20)
\DashLine(100,50)(140,50){1.2}
\end{picture}
\begin{picture}(200,100)(0,0)
\SetWidth{1.8} \Gluon(20,20)(60,20){3}{5}
\Gluon(20,80)(60,80){3}{5} \ArrowLine(60,80)(60,20)
\ArrowLine(100,50)(60,80) \ArrowLine(60,20)(100,50)
\DashLine(100,50)(140,50){1.2}
\end{picture} \\
\caption{\small Feynman diagram for
 $g(p_1)+g(p_2)\rightarrow \tilde{\nu}(q)$ via quark loops.}
 \label{figgg}
\end{center}
\end{figure}
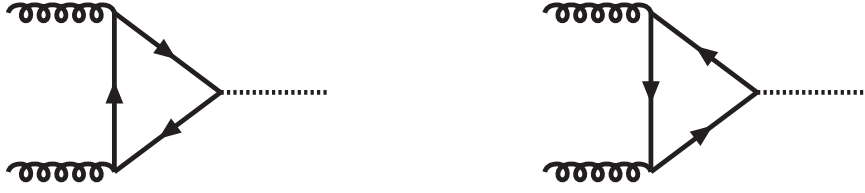

This process comes from diagrams shown  in Fig.~\ref{figgg}, and the
corresponding amplitude is
\begin{eqnarray}
\label{eqc}
{\cal M}&=& -{\lambda'} \, g_s^2 \, \frac{\delta_{ab}}{2} \, \epsilon^{\mu}(p_1) \,
         \epsilon^{\nu}(p_2) \, {M}_{\mu\nu}, \\
{M}_{\mu\nu}&=&\int \frac{d^4 l}{(2\pi)^4}
\frac{N_{\mu\nu}}{[l^2-m_q^2+i\epsilon]
[(l+p_1)^2-m_q^2+i \epsilon] [(l+p_1+p_2)^2-m_q^2+i \epsilon]},
\nonumber \\
N_{\mu\nu} &=& Tr\Big\{\gamma_{\mu}(\not{l}+m_q)
\frac{1+\gamma_5}{2} (\not{l}+\not{p}_1+\not{p}_2+m_q]
\gamma_{\nu} [\not{l}+\not{p}_1+m_q]\Big\} \nonumber \\
 &&-Tr\Big\{ \gamma_{\mu} (\not{l}+\not{p}_1-m_q)\gamma_{\nu}
[\not{l}+\not{p}_1+\not{p}_2-m_q]\frac{1+\gamma_5}{2}(\not{l}-m_q)\Big\}\;.
\nonumber
\end{eqnarray}
With  a straightforward calculation, we obtain the cross section of
the subprocess:
\begin{eqnarray}
\hat{\sigma}_{gg}&=&\frac{1}{2\hat{s}}
\frac{1}{4(N_c^2-1)^2} \int |{\cal M}|^2 \frac{d^3 q}{(2\pi)^3 2q^0}
(2\pi)^4 \delta^4 (p_1+p_2-q) \\
&=&\frac{\alpha_s^2 {\lambda'}^2}{16\pi \, (N_c^2-1)} \frac{m_q^2}{\hat{s}^2} \,
\Big\{ [(\hat{s}-2m_q^2)^2+4m_q^4]|C_0|^2+2 + (8 m^2_q-2\hat{s}) Re[C_0] \Big\}
~\delta(1-\tau),
\nonumber
\end{eqnarray}
where $C_0$ is the loop integral and is given by
\begin{eqnarray}
C_0&=&\int \frac{d^4 l}{i\pi^2} \frac{1}{(l^2-m_q^2)\, [(l+p_1)^2-m_q^2]
\, [(l+p_1+p_2)^2-m_q^2]} \nonumber  \\
&=&\frac{1}{\hat{s}} \int\limits_0^1 \frac{d z}{z}
\ln\,\Big(\,1-\frac{\hat{s}}{m_q^2} z+\frac{\hat{s}}{m_q^2}
z^2\,\Big)\;.
\end{eqnarray}
The imaginary part arises when crossing the threshold
$\hat{s}=4m_q^2$. In this case, $C_0$ reads
\begin{equation}
C_0=\frac{1}{\hat{s}}\,\Big[\; \frac{1}{2} \,
\left(\,\ln^2\omega-\pi^2\,\right)\, +\, i~\pi\, \ln \omega \;
\Big] \Theta(m_{\tilde{l}}-2m_q)\;,
\end{equation}
with
\begin{equation}
\omega=\frac{1-\sqrt{1-\frac{4m_q^2}{\hat{s}}}}
{1+\sqrt{1-\frac{4m_q^2}{\hat{s}}}}\;.
\end{equation}
We see that the cross section $\hat{\sigma}_{gg}$ is approximately
proportional to the quark mass squared in the loops. Thus we expect
that only the $b$-quark loops give the most important contributions.

\begin{figure}[tb]
\centerline{ 
\psfig{figure=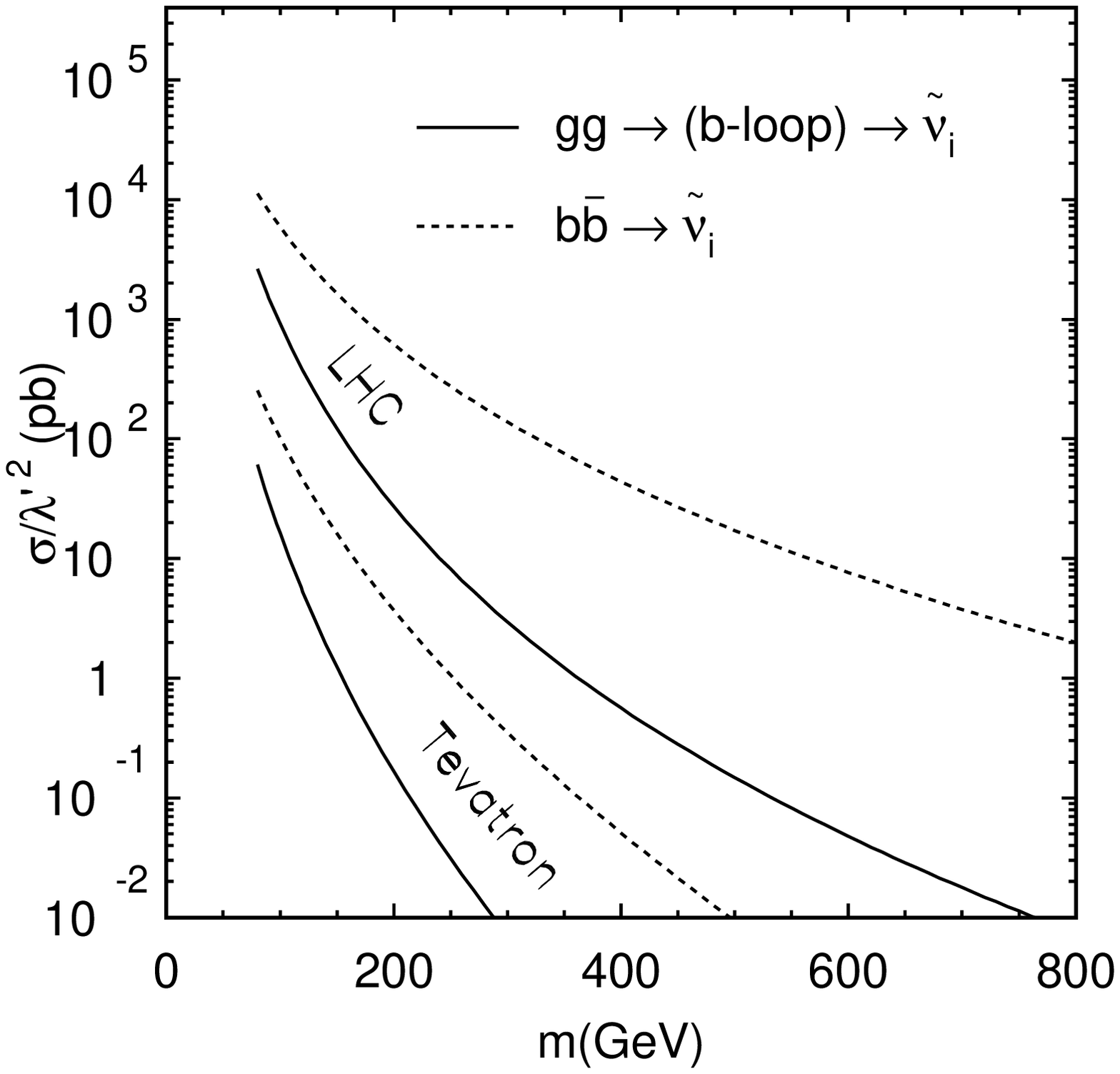,width=8cm}}
\caption{Cross section related to ${\lambda'}_{i33}$ for 
$b\bar{b}\rightarrow \tilde{\nu}_i+X$ 
and  $gg \rightarrow b-loop \rightarrow
\tilde{\nu}_i$ at Tevatron ($\sqrt{s}=2$ TeV) and
LHC ($\sqrt{s}=14$ TeV).}\label{csl33}
\end{figure}

Given the cross section of the subprocess,  the cross section at
hadron colliders can also be calculated using the QCD
factorization formula (\ref{QCD-fac}).  The results at LHC and
Tevatron shown in Fig.~\ref{csl33} are obtained by using
CTEQ6L. In our calculation, we keep only the contribution from the
$b$-quark loops and set $b$-quark mass to be 5 GeV. 
The subprocess cross sections for the loop-induced diagrams
scale as $\lambda^2/m_{\tilde{\nu}}^2$.
The total cross sections after folding in the parton distribution functions
decrease even faster versus $m_{\tilde{\nu}}^2$.  For comparison,
we show the results for $b\bar{b}\rightarrow \tilde{\nu}_i+X$ in
Fig.~\ref{csl33}, again. Obviously, the production rate of
$gg\rightarrow b{\rm -loop} \rightarrow \tilde{\nu}_i$ is much smaller
than that of $b\bar{b}\rightarrow \tilde{\nu}_i$ (for details, see Table \ref{tabgg}).

\begin{table}[tb]
\begin{center}
\begin{tabular}{|l|c|cll|} \hline
$\lambda'$         & Bounds  &\multicolumn{3}{c|}{$\sigma _{m=150-500\, GeV}$} \\
                   &         & &  Tevatron(Pb)      & LHC(Pb)  \\ \hline
$\lambda'_{133} $  & 0.18    &$b\bar{b}\rightarrow \tilde{\nu}_e$  &
0.52--3$\cdot 10^{-4}$   & 53--0.56   \\ \cline{3-5}
                   &         &$gg\rightarrow \tilde{\nu}_e$    &0.04--2$\cdot 10^{-6}$  & 4.0--5$\cdot 10^{-3}$ \\ \hline\hline
$\lambda'_{233} $  & 0.45    &$b\bar{b}\rightarrow\tilde{\nu}_{\mu}$
& 3.3--2$\cdot 10^{-3}$ & 333--3.5   \\\cline{3-5}
                   &         &$gg\rightarrow \tilde{\nu}_{\mu}$    & 0.25--$10^{-5}$ & 25--0.03 \\ \hline \hline
$\lambda'_{333} $  & 0.58 &$b\bar{b}\rightarrow\tilde{\nu}_{\tau}$ &
5.4--3$\cdot 10^{-3}$ & 553--5.8   \\ \cline{3-5}
                   &         &$gg\rightarrow \tilde{\nu}_{\tau}$    &0.42--2$\cdot 10^{-5}$  & 41--0.05 \\ \hline
\end{tabular}
\end{center}
\caption{The results for $b\bar{b}\rightarrow \tilde{\nu}+X$ and $gg\rightarrow \tilde{\nu}$.}\label{tabgg}
\end{table}

\section{Summary}
\label{sec6}

We have evaluated the cross section for single slepton production at hadron
colliders in supersymmetric theories with $R$-parity violating interactions
to the next-to-leading order in QCD. 
At the LHC, the QCD corrections are rather stable and 
increasing monotonically versus $m_{\tilde\ell}$. 
The $K$-factors are typically around $1.2-1.4$, except that the
range becomes slightly larger when more sea quarks are involved.
At the Tevatron, the $K$-factor is around $1.1-1.5$ when 
the valence quarks dominate. 
For other non-valence quark channels, 
The $K$-factors  can be larger than a factor of two at large $x$-values.
The event rates are sizable and the assumption for the Majorana mass
generation via those $R$-parity violating interactions 
may be tested at the hadron collider experiments.

We obtained fully differential cross section and 
performed soft-gluon resummation to all
order  in $\alpha_s$ of  leading logarithm to obtain a complete 
transverse momentum spectrum of the slepton. 
We find that  the full transverse momentum 
spectrum is peaked at a few  GeV, consistent with the early
results for Drell-Yan production of lepton pairs.

We also calculated the contribution from gluon fusion
via quark-triangle loop diagrams. 
The cross section of this process is significantly smaller
than that of the  tree-level process induced by the initial $b\bar{b}$
annihilation.

\bigskip                                                                                 
{\bf Note added:}
 After completing our studies, we were made aware of a recent
publication that dealt with similar topics in Ref.~\cite{Dreiner:2006sv}.
For the parts we overlapped, 
namely the SM QCD corrections and soft gluon resummation, our results
are in agreement with each other.

\bigskip
\noindent{\bf Acknowledgments} This work was supported in part by
National Natural Science Foundation of China (NNSFC). The work of
T.H.  was supported in part by a DOE grant No. DE-FG02-95ER40896,
by Wisconsin Alumni Research Found. The work of Z.G.S  was
supported in part by NCET of MoE China and Huo Yingdong Foundation.



\begin{thebibliography}{999}

\bibitem{Nureview} For recent reviews on neutrino masses and
flavor oscillations, see {\it e.g.},
B.~Kayser, p.~145 of the Review of Particle Physics,
Phys.~Lett.~{\bf B592}, 1 (2004);
V.~Barger, D.~Marfatia, and K.~Whisnant,
Int.~J.~Mod.~Phys.~{\bf E12}, 569 (2003).

\bibitem{Rmass}
C. S. Aulakh and R. N. Mohapatra,  Phys. Lett. {\bf B119}, 136 (1982);  
%
  L.~J.~Hall and M.~Suzuki,
  Nucl.\ Phys.\ B {\bf 231}, 419 (1984);
  S.~Dawson,
  Nucl.\ Phys.\ B {\bf 261}, 297 (1985); 
  J.~R.~Ellis, G.~Gelmini, C.~Jarlskog, G.~G.~Ross and J.~W.~F.~Valle,
  Phys.\ Lett.\ B {\bf 150}, 142 (1985);
   G. G. Ross and J. W. F. Valle, Phys. Lett. {\bf B151}, 375 (1985).
  
\bibitem{Rnew}
  M.~Drees, S.~Pakvasa, X.~Tata and T.~ter Veldhuis,
  Phys.\ Rev.\ D {\bf 57}, 5335 (1998)  [arXiv:hep-ph/9712392]; 
  E. J. Chun, S. K. Kang, C. W. Kim and U. W. Lee, Nucl. Phys. {\bf B544}, 
  89 (1999)  [{arXiv:hep-ph/9807327}]; 
%
  V.~D.~Barger, T.~Han, S.~Hesselbach and D.~Marfatia,
  Phys.\ Lett.\ B {\bf 538}, 346 (2002)
  [arXiv:hep-ph/0108261].

\bibitem{Rold}
  S.~Dimopoulos and L.~J.~Hall,
  Phys.\ Lett.\ B {\bf 207} (1988) 210;
  %
 V.~D.~Barger, G.~F.~Giudice and T.~Han,
  Phys.\ Rev.\ D {\bf 40}, 2987 (1989);
 H.~K.~Dreiner and R.~J.~N.~Phillips,
  Nucl.\ Phys.\ B {\bf 367}, 591 (1991).

\bibitem{tevrp}
  V.~M.~Abazov {\it et al.}  [D0 Collaboration],
  Phys.\ Rev.\ Lett.\  {\bf 97}, 111801 (2006);
%
A.~Abulencia {\it et al.}  [CDF Collaboration],
  Phys.\ Rev.\ Lett.\  {\bf 96}, 211802 (2006).

\bibitem{Barbier04ez}
For a recent review on $R$-parity violation, see {\it e. g.}, 
  R.~Barbier {\it et al.},
  Phys.\ Rept.\  {\bf 420}, 1 (2005)  [arXiv:hep-ph/0406039].

\bibitem{Choudhury2003}
D. Choudhury, S. Majhi, V. Ravindran, Nucl. Phys. B{\bf 660}, 343 (2003).

\bibitem{Yang:2005ts}
  L.~L.~Yang, C.~S.~Li, J.~J.~Liu and Q.~Li,
  Phys.\ Rev.\ D {\bf 72}, 074026 (2005)
  [arXiv:hep-ph/0507331].

\bibitem{Harris01sx}
 H.~Baer, J.~Ohnemus and J.~F.~Owens,
  Phys.\ Lett.\ B {\bf 234}, 127 (1990);
  B.~W.~Harris and J.~F.~Owens,
  Phys.\ Rev.\ D {\bf 65}, 094032 (2002)
  [arXiv:hep-ph/0102128].

\bibitem{Pumplin02vw}
  J.~Pumplin, D.~R.~Stump, J.~Huston, H.~L.~Lai, P.~Nadolsky and W.~K.~Tung,
  JHEP {\bf 0207} (2002) 012
  [arXiv:hep-ph/0201195].


%
\bibitem{Han:1991sa}
  J.~C.~Collins, D.~E.~Soper and G.~Sterman,
  Nucl.\ Phys.\ B {\bf 223}, 381 (1983);
P.~B.~Arnold and R.~P.~Kauffman,
Nucl.\ Phys.\ B {\bf 349}, 381 (1991);
  T.~Han, R.~Meng and J.~Ohnemus,
  Nucl.\ Phys.\ B {\bf 384}, 59 (1992).
  


%
%
%
%
%
%
%
%
%
%
%
%
%
%
%
\bibitem{cpy}
  F.~Landry, R.~Brock, G.~Ladinsky and C.~P.~Yuan,
  Phys.\ Rev.\  D {\bf 63}, 013004 (2001)
  [arXiv:hep-ph/9905391].

\bibitem{Dreiner:2006sv}
  H.~K.~Dreiner, S.~Grab, M.~Kramer and M.~K.~Trenkel,
  arXiv:hep-ph/0611195.

\end{thebibliography}
\end{document}